\documentclass[a4paper]{article}

\usepackage[english]{babel}
\usepackage[utf8x]{inputenc}
\usepackage[T1]{fontenc}

\usepackage[a4paper,top=3cm,bottom=2cm,left=3cm,right=3cm,marginparwidth=1.75cm]{geometry}

\usepackage{amsmath}
\usepackage{graphicx}
\usepackage[colorinlistoftodos]{todonotes}
\usepackage[colorlinks=true, allcolors=blue]{hyperref}
\usepackage{comment}
\usepackage{bm}
\usepackage{graphicx,psfrag,epsfig,subfig,comment,array}
\usepackage{amsmath,amssymb,mathtools}
\usepackage{latexsym}
\usepackage{multicol,xspace}
\usepackage{float}
\usepackage{booktabs}
\usepackage{multirow}
\usepackage{rotating}
\usepackage{bm}
\usepackage{tabularx}
\usepackage{amsthm}
\usepackage{flushend}
\usepackage{algorithm,algpseudocode}
\usepackage{pbox,footnote}

\newtheorem{proposition}{Proposition}
\newtheorem{theorem}{Theorem}

\title{Maximum Correntropy Derivative-Free Robust Kalman Filter and Smoother}
\author{Hongwei Wang\footnote{H. Wang is with the School of Aeronautics, Northwestern Polytechnical University, Xian 710072, China, and also with the Department of Electrical and Computer Engineering, Stevens Institute of Technology, Hoboken, NJ 07030 USA (e-mail: tianhangxinxiang@163.com).}, Hongbin Li\footnote{H. Li is with the Department of Electrical and Computer Engineering, Stevens Institute of Technology, Hoboken, NJ 07030 USA (e-mail: hli@stevens.edu).}, Junyi Zuo\footnote{J. Zuo is with the School of Aeronautics, Northwestern Polytechnical University, Xian 710072, China (e-mail: junyizuo@nwpu.edu.cn).}, Wei Zhang\footnote{W. Zhang is with the School of Aeronautics, Northwestern Polytechnical University, Xian 710072, China (e-mail: weizhangxian@nwpu.edu.cn).} and Heping Wang\footnote{H. Wang is with the School of Aeronautics, Northwestern Polytechnical University, Xian 710072, China (e-mail: wangheping@nwpu.edu.cn).}}

\begin{document}
\maketitle

\begin{abstract}
We consider the problem of robust estimation involving filtering and smoothing for nonlinear state space models which are disturbed by heavy-tailed impulsive noises. To deal with heavy-tailed noises and improve the robustness of the traditional nonlinear Gaussian Kalman filter and smoother, we propose in this work a general framework of robust filtering and smoothing, which adopts a new maximum correntropy criterion to replace the minimum mean square error for state estimation. To facilitate understanding, we present our robust framework in conjunction with the cubature Kalman filter and smoother. A half-quadratic optimization method is utilized to solve the formulated robust estimation problems, which leads to a new maximum correntropy derivative-free robust Kalman filter and smoother. Simulation results show that the proposed methods achieve a substantial performance improvement over the conventional and existing robust ones with slight computational time increase.
\end{abstract}

\section{Introduction}

In recent years, estimation problems involving filtering and smoothing based on dynamic state space models (SSMs) have received significant attentions. These problems are frequently encountered in many areas, such as target tracking, fault detection and diagnosis, parameter estimation, navigation, and many others. The celebrated Kalman filter (KF)~\cite{Kalman:1960} offers optimal estimation with the minimum mean square error (MMSE) for a linear SSM when both the process and measurement noises are Gaussian. Nevertheless, most applications in practice inherently have nonlinear SSMs, for which an optimal nonlinear filter or smoother is typically intractable. Several sub-optimal solutions were proposed for nonlinear SSMs, including the extended KF~\cite{anderson:1979}, unscented KF~\cite{Julier:2000}, cubature Kalman filter (CKF)~\cite{arasaratnam:2009}, and others. Some interesting relations among these solutions can be found  in~\cite{gustafsson:2012,wu:2006}. Meanwhile, for smoothing, the Rauch-Tung-Striebel (RTS) smoother was introduced in~\cite{RTS:1965} for linear Gaussian SSMs. A general framework of Gaussian optimal smoothing for nonlinear SSMs was proposed in~\cite{sarkka:2010}, following which several sub-optimal nonlinear RTS smoothers were developed, such as the extended RTS smoother, unscented RTS smoother, cubature Kalman smoother (CKS), and others.

Although the aforementioned filtering and smoothing methods in general perform well under the Gaussian assumption, they can potentially break down in the presence of heavy-tailed non-Gaussian noises, which may appear in either the process procedure or measurement procedure. This happens in, for instance, tracking a maneuvering target with outliers in observations~\cite{bilik:2010}. A primary reason behind the degradation is that the traditional KF/RTS and their sub-optimal extensions were derived from the MMSE criterion, and the resulting estimates are sensitive to heavy-tailed noises~\cite{schick:1994}.

Persistent efforts have been devoted to tackling heavy-tailed noises in order to obtain more robust state estimates. Multiple models based techniques and sequential Monte Carlo sampling methods can be used to handle non-Gaussian noises~\cite{kotecha:2003,arulampalam:2002}. Unfortunately, their heavy computational burden makes them considerably more difficult to implement in real time. Another common strategy to enhance the robustness is robust statistics (or influence functions). Masreliez and Martin~\cite{masreliez:1977} introduced the Huber cost function~\cite{Huber:2009} to a linear SSM to construct a robust KF by recasting the filtering problem as a linear regression. After that, this type of robust filters were extended to nonlinear SSMs via linearization~\cite{Gandhi:2010}. More recently, to reduce the approximation error arising from linearization, nonlinear regression based Huber's robust filters were proposed in~\cite{karlgaard:2015,Chang:2017}. Other approaches for robust estimation, e.g., the $H_{\infty}$ filter~\cite{hu2011second} and heavy-tailed distributions based filters and smoothers~\cite{Huang:2016,agamennoni:2012,huang:2016_2}, were also studied.

In addition, several recent studies explored using optimization to design robust smoothers since smoothing is in essence an optimization problem. Specifically, an $\ell_1$ robust Kalman smoother was presented in~\cite{aravkin2011ell} based on an interior point method, where the Laplace distribution was utilized to model the measurement noise. Quadratic programming was employed to design a student's t Kalman smoother in~\cite{aravkin2014robust}, where both the process and measurement noise were modeled by student's t distribution. Quadratic support functions were introduced in~\cite{aravkin2013sparse} to present generalized Kalman smoothing methods. Other optimization methods for robust Kalman smoothing were discussed in ~\cite{aravkin:2017generalized}. Although optimization based methods provide reasonable smoothing results, they still have some limitations. A major one is that the methods ignore covariance propagation, and hence some important statistic information about the state may get lost. Besides, these methods were primarily designed for smoothing, and it is difficult to extend them for filtering, which is more frequently encountered in real-time applications.

Recently a concept called correntropy in information theoretic learning and an associated maximum correntropy criterion (MCC)~\cite{Liu:2007} was utilized as a new cost function for robust filtering~\cite{liu:2017}. As shown in~\cite{Liu:2007}, MCC has a close relationship with M-estimators. In fact, correntropy is a robust recasting of the Welsch cost function~\cite{dennis:1978}. Compared with Huber's cost function, the Welsch function penalizes a large fitting errors~\cite{Liu:2007}. Furthermore, the Welsch M-estimator outperforms Huber's M-estimator in dealing with different types of outliers. Meanwhile the kernel size in the Welsch is less sensitive to select than the threshold in Huber's M-estimator, which can have a significant impact on the estimation performance~\cite{He:2014}.

In the present paper, we focus on developing a general robust framework, including both robust filtering and robust smoothing, to estimate the states of nonlinear SSMs where heavy-tailed noises may be present in both the process and measurement procedures. The proposed framework is based on the MCC and a half-quadratic (HQ) optimization method. Numerical results show that the proposed algorithms can provide reliable state estimation, and are robust to almost all levels of contamination. In comparison with existing robust algorithms, our methods yield more accurate results with a moderate increase of computation time.

\subsection{Related Work}

Correntropy and its associated MCC criterion was originally proposed in~\cite{Liu:2007}, where the relationship between MCC and M-estimation was established. The advantages of the MCC in dealing with non-Gaussian noises with heavy tails were demonstrated in~\cite{Liu:2007} via numerical simulations. Thereafter, the MCC criterion was employed to improve the robustness in several areas, such as feature selection~\cite{Yan:2011}, robust face recognition~\cite{he:2011maximum}, principal component analysis~\cite{he:2011_2}, and so on. During last few years, several researchers started to employ the MCC to deal with heavy-tailed noises in Kalman filtering. It was first utilized to improve the robustness of the KF in~\cite{Cinar:2011,Cinar:2012}, where gradient descent was used for state estimation. Those methods, however, ignored covariance propagation, which is an important part in the KF.

In~\cite{izanloo:2016}, a robust KF called MCC-KF was proposed, which integrated both the MCC and weighted least squares into the traditional KF framework. A square-root format of the MCC-KF was proposed in~\cite{kulikova:2017} to improve the numerical stability. MCC-KF and its square-root version only modified the weights of the measurements and thus had the same prediction step as the traditional KF which was derived under the Gaussian assumption. Consequently, the MCC-KF and its square-root version were sensitive to non-Gaussian process noise. To handle this issue, another robust KF called MCKF~\cite{chen:2017} was developed using an approach similar to that of the Huber based robust KF, i.e., recasting the filtering problem as a linear regression. In MCKF, a fixed-point algorithm was employed to iteratively provide robust estimation, and both the prediction step and measurement update were adapted. Henceforth, several variants of the MCKF were proposed for nonlinear systems by linearizing the nonlinear mappings, e.g., the Taylor series~\cite{liu2:2016}, unscented transformation~\cite{liu:2017}, and Gauss-Hermite quadrature rule~\cite{qin:2017}. Most recently, the MCC was employed to solve the robust estimation problem in the continuous-discrete system~\cite{kulikov:2018}. In addition, the MCC was also utilized to suppress outliers in measurements assuming Gaussian noise in the process~\cite{wang:2016,wang:2017}.

Although above studies have clearly demonstrated the potential of the MCC for robust Kalman filtering, there are some deficiencies with the state-of-the-art research in this area. Specifically, a general framework of using MCC for robust Kalman filtering and smoothing is missing.  Most existing works have addressed the problem under a specific scenario, but the resulting methods may degrade significantly under a different scenario. For example, the method in \cite{liu:2017} was developed under the assumption that the process and measurement are both contaminated, its performance may experience considerable degradation under a different scenario, e.g., when only the process or the measurement is contaminated. In addition, most recently developed robust Kalman filtering schemes using the MCC were based on linear regression~\cite{chen:2017,liu2:2016,liu:2017,qin:2017}, where a linearization procedure is inevitable, resulting in a loss of accuracy. Another limitation is that no existing work has explored MCC for robust Kalman smoothing.

\subsection{Main Contributions}

The main contributions of this paper are summarized as follows.
\begin{itemize}
  \item We propose a general framework that integrates MCC for robust Kalman filtering. We take an optimization perspective based approach, rather than linear regression as employed in most existing MCC based robust Kalman filtering solutions. This approach bypasses the linearization procedure, which not only makes our methods derivative-free but also reduces approximation errors caused by linearization.
  \item A distinctive feature of the proposed framework involves replacing the quadratic cost in the traditional KF by a sum of weighted instantaneous correntropy terms to improve robustness. The weight of each correntropy is automatically tuned to cope with different scenarios involving Gaussian or heavy-tailed non-Gaussian noises.
  \item The optimization problem underlying the proposed framework is solved by a half-quadratic method. This enables our robust Kalman filtering algorithm to be cast using the conventional KF iterative procedures, thus inheriting benefits associated with the latter.
  \item We also extend the above robust filtering framework for robust smoothing, resulting in a MCC based robust Kalman smoother. To the best of our knowledge, our approach is the first robust nonlinear Kalman smoother based on the MCC.
\end{itemize}

The paper is organized as follows. We first formulate the problem in Section~\ref{Set:prom}, and derive the robust nonlinear Kalman filter based on MCC in Section~\ref{Sec:fmcc}. In Section~\ref{smcc}, the robust nonlinear Kalman smoother is presented. We show numerical examples to illustrate the performance of the proposed robust filtering and smoothing algorithms in Section~\ref{simu}. Finally, we draw conclusions and summarize future related research in Section~\ref{con}.

\emph{Notations:} In this paper, normal font letters are scalars, boldface lowercase letters denote column vectors, and boldface uppercase letters mean matrices. The estimate of state $\bm x_t$ given the measurements from $t=1$ to $t=m$ is denoted by $\hat{\bm x}_{t|m}$ where $m=t$ means filtering, and $m=T$ means smoothing, with $T$ denoting the total number of observations. $\bm I_n$ denotes the $n$-dimensional identity matrix. $\mathcal{N}(\cdot,\cdot)$ denotes a Gaussian distribution.

\section{Problem Formulation}
\label{Set:prom}

In this paper, we address a sequential estimation problem in nonlinear systems involving non-Gaussian noises. More specifically, both the process noise and measurement noise are primarily Gaussian random processes; but may occasionally be contaminated by outliers with an unknown distribution. To model this, we consider the following nonlinear discrete-time state-space model (SSM)
\begin{align}
\bm x_t &= \bm f(\bm x_{t-1})+(1-\rho_1)\bm v_{t-1}+\rho_1\bm\varepsilon_1\label{SSM1}\\
\bm y_t &= \bm h(\bm x_t) +(1-\rho_2) \bm w_t+\rho_2\bm\varepsilon_2\label{SSM2}
\end{align}
where $t=\{1,\cdots,T\}$ is the time index; $\bm x_t\in\mathbb{R}^n$ is the state of interest to be estimated; $\bm y_t\in\mathbb{R}^m$ is the observation; $\bm f(\cdot)$ and $\bm h(\cdot)$ are some known nonlinear mappings, representing the state transition and measurement procedure, respectively; $\bm v_{t-1}$ and $\bm w_t$ are zero-mean Gaussian distribution vectors with $\bm Q_{t-1}$ and $\bm R_t$ as covariance, representing the nominal process noise and observation noise, respectively; $\bm\varepsilon_1$ and $\bm\varepsilon_2$ are some contaminating noises; $\rho_1$ and $\rho_2$ are the corresponding contamination ratios. The initial state ${\bm x}_{0}$ is assumed to follow a known Gaussian prior distribution $p(\bm x_0)=\mathcal{N}(\hat{{\bm x}}_{0|0},\bm P_{0|0})$. Furthermore, $\bm x_0$, $\bm v_{t-1}$, $\bm w_t$, $\bm\varepsilon_1$ and $\bm \varepsilon_2$ are assumed to be mutually independent.

Practically, the distributions of the contaminating noises and their corresponding contamination ratios are often unknown. If one chooses to ignore the contaminating noises and run a Gaussian approximation filter and smoother, the performance will likely to be unsatisfying or even diverge~\cite{Huber:2009}. Although ineffective in dealing with the contaminating noise, the Gaussian approximation filter and smoother are popular in those nonlinear sequential estimation when both process and measurement noises are Gaussian due to their computational efficiency and relative high accuracy. This motivates us to explore extended versions of the Gaussian approximation filter and smoother to deal with the contaminating noise in nonlinear systems.

The aim of this work is to develop a unified framework for robust nonlinear filter and smoother design with the information $p(\bm x_0)$, $p(\bm v_t)$, $p(\bm w_t)$ and the observations (namely, $\bm y_{1:t}$ for filtering and $\bm y_{1:T}$ for smoothing, respectively). To address non-Gaussian contaminating noise, the proposed framework employs a new maximum correntropy criterion (MCC)~\cite{Liu:2007} which integrated with the Gaussian approximation filter and smoother, resulting in a class of MCC based filter and smoother solutions. The developed robust filtering and smoothing techniques are expected to have a similar performance as the Gaussian approximation counterparts in the absence of outliers, meanwhile exhibiting less sensitivity in scenarios involving contaminating noise.

In the following, our robust framework is presented in conjunction with the CKF~\cite{arasaratnam:2009} and CKS~\cite{arasaratnam:2011}, which are popular for nonlinear system estimation. A brief summary of the CKF and CKS is included in Appendix \ref{A} to facilitate the presentation of our framework. It is straightforward to extend the framework for use with other nonlinear Gaussian approximation filters and smoothers.

\section{Robust Nonlinear Kalman filter with MCC}
\label{Sec:fmcc}
\subsection{MCC-Based Filtering}
\label{par_dis}

In the Bayesian filtering paradigm, the core problem is to calculate the posterior density through Bayes' rule based on the measurements up to $t$,
\begin{align}
p(\bm x_t|\bm y_{1:t})\propto{p(\bm y_t|\bm x_t)p(\bm x_t|\bm y_{1:t-1})}
\label{meaupt}
\end{align}
where $p(\bm y_t|\bm x_t)$ is the likelihood function which is defined by the measurement process and $p(\bm x_k|\bm y_{1:t-1})$ is the predictive density which is controlled by Chapman-Kolmogorov equation
\begin{align}
p(\bm x_t|\bm y_{1:t-1})=\int{p(\bm x_t|\bm x_{t-1})p(\bm x_{t-1}|\bm y_{1:t-1})d{\bm x_{t-1}}}
\label{timeupt}
\end{align}
where $p(\bm x_{t-1}|\bm y_{1:t-1})$ is the posterior density at time $t-1$ and $p(\bm x_t|\bm x_{t-1})$ is the transition density which is defined by the state transition process. The maximum a posteriori estimation (MAP) of $\bm x_t$ can be obtained through minimizing the negative log posterior density, i.e.,
\begin{align}
\hat{\bm x}_{t|t}=&\text{arg}\min_{\bm x_t}-\log p(\bm x_t|\bm y_{1:t})\notag\\
=&\text{arg}\min_{\bm x_t}\Big(-\log p(\bm x_t|\bm y_{1:t-1})-\log p(\bm y_t|\bm x_t)\Big)
\label{cost1}
\end{align}
Under the Gaussian approximation, i.e., both the process noise and measurement noise are assumed Gaussian and the predictive density is a Gaussian distribution, dropping terms that do not depend on $\bm x_t$, we can rewrite (\ref{cost1}) as~\cite{Chang:2017}
\begin{align}
\hat{\bm x}_{t|t}=\text{arg}\min_{\bm x_t}\Big(\frac{1}{2}\|\bm x_t-\hat{\bm x}_{t|t-1}\|^2_{{\bm P}_{t|t-1}^{-1}}
+\frac{1}{2}\|\bm y_t-h(\bm x_t)\|^2_{{\bm R}_{t}^{-1}}\Big)
\label{cost2}
\end{align}
where $\|\bm z\|_{\bm A}^2=\bm z^T\bm A\bm z$ and $\hat{\bm x}_{t|t-1}$ is the predicted state with the prediction error covariance $\bm P_{t|t-1}$. We define the normalized error as
\begin{align}
\bm \alpha_{t}&=\bm P_{t|t-1}^{-1/2}\left(\bm x_t-\hat{\bm x}_{t|t-1}\right)\label{def1} \\
\bm \beta_{t}&=\bm R_{t}^{-1/2}\left(\bm y_t-h(\bm x_t)\right)\label{def2}
\end{align}
Then (\ref{cost2}) can be further interpreted as
\begin{align}
\hat{\bm x}_{t|t}=&\text{arg}\min_{\bm x_t}\ \left(\frac{1}{2}\bm \alpha_{t}^T\bm \alpha_{t}+\frac{1}{2}\bm \beta_{t}^T\bm \beta_{t}\right)\notag\\
=&\text{arg}\min_{\bm x_t}\ \left(\frac{1}{2}\sum_{i=1}^n\alpha_{t,i}^2
+\frac{1}{2}\sum_{j=1}^m\beta_{t,j}^2\right)
\label{cost:2}
\end{align}
where $\alpha_{t,i}$ is the $i$-th component of $\bm \alpha_t$ and $\beta_{t,j}$ is the $j$-th component of $\bm \beta_t$.

Equation \eqref{cost:2} shows that the Gaussian approximation filtering approach is actually a minimum mean square error (MMSE) estimator. It is clear that the quadratic function (i.e., $\ell_2$ norm) in MMSE has an apparent effect of emphasizing the contribution of large errors, which makes the MMSE estimator sensitive to impulsive noise. To improve the robustness, the $\ell_2$-norm objective function should be replaced by another cost function which is insensitive to the impulsive noise.

We propose to integrate the concept of correntropy~\cite{Liu:2007}, which has been shown effective in dealing with outliers~\cite{mandanas:2017}, for robust Kalman filtering and smoothing, by utilizing the maximum correntropy criterion (MCC) to replace the MMSE criterion. Specifically, we apply the instantaneous correntropy to each component of the normalized error, resulting in the following cost function for robust nonlinear filtering:
\begin{align}
\hat{\bm x}_{t|t}=\text{arg}\max_{\bm x_t}\left(\sum_{i=1}^n a_{t,i}\kappa_{\sigma_{t,i}}(\alpha_{t,i})
+\sum_{j=1}^m b_{t,j}\kappa_{\eta_{t,j}}(\beta_{t,j})\right)
\label{cost_ro}
\end{align}
where $\kappa_\sigma(\cdot)$ is the kernel function which satisfies Mercer's Theorem~\cite{Vapnik:1995}; $a_{t,i}$ and $b_{t,j}$ are weighting coefficients to be further determined. The kernel function plays a central role in MCC and different kernel functions provide different estimation results.

Without loss of generality,  we apply the the Gaussian kernel function with bandwidth $\sigma$,
\begin{align*}
\kappa_{\sigma}(e)=\exp\left({-\frac{e^2}{2\sigma^2}}\right).
\end{align*}
The Gaussian kernel based MCC is closely related to the Welsch M-estimator, where outliers are given small weights in optimization~\cite{He:2014}. For completeness, in Appendix \ref{B}, we provide a short review of the correntropy, MCC, and related properities.

There are two aspects that should be noted in \eqref{cost_ro}. First, a weighting coefficient (namely $a_{t,i}$ and $b_{t,j}$) is introduced to each correntropy. These coefficients should be properly selected to maintain consistency to the Gaussian approximation filter when the SSM does not involve impulse noises (see Section \ref{sle_par} for details on selection for these parameters). Second, having distinct bandwidth for each correntropy offers additional flexibility of adaptiveness to different applications. One possible strategy is to use the same kernel bandwidth for each component of the predictive error and a different kernel bandwidth for the measurement error terms in (\ref{cost_ro}). This makes the proposed algorithms suitable for handling different process noise and measurement noise. For example, in a case where outliers only occur in the measurement procedure, we can select a large $\sigma_t$ so that the resulting MCC behaves like a quadratic criterion, and a smaller $\eta_t$ to defend the outliers.

\subsection{A Half-Quadratic Optimization Based Solution}

\label{sle_par}

A number of optimization techniques such as the steepest descent~\cite{Liu:2007} and the fixed point iteration approach~\cite{Chen:2015} can in principle be applied to solve MCC based problems. However, our correntropy criterion~\eqref{cost_ro} involves different kernel bandwidths. The problem can be more efficiently solved by a half-quadratic approach~\cite{Geman:1992}. Specifically, our proposed solution to \eqref{cost_ro} applies the half-quadratic approach in an iterative manner, through which auxiliary variables are introduced to transform the original problem into a quadratic problem. The resulting quadratic problem can be cast in the conventional Gaussian approximation filtering framework. This is a major advantage of our proposed solution. The half-quadratic approach is based on the concept of the convex conjugate function and, in particular, the following result.

\begin{proposition}
\label{propo}
There exists a convex function $\psi$: $[-1,0)$ $\to$ $ [-1,0)$, such that
\begin{align}
\kappa_{\sigma}(x)=\sup_{-1\le p<0}\left(\frac{p}{2\sigma^2}x^2
-\psi(p)\right)
\label{pro1}
\end{align}
and for a fixed $x$, the supremum is reached at $p=-\kappa_{\sigma}(x)$.

Proof: see Appendix~\ref{prop_proof}.
\end{proposition}

Using (\ref{pro1}), the robust filtering problem \eqref{cost_ro} can be reinterpreted in an augmented format as
\begin{align}
\hat{\bm x}_{t|t}=\text{arg}\max_{\bm p_t,\bm q_t,\bm x_t}\
\left\{\sum_{i=1}^n a_{t,i}\left(\frac{p_{t,i}}{2\sigma_{t,i}^2}\alpha_{t,i}^2
-\psi(p_{t,i})\right)
+\sum_{j=1}^m b_{t,j}\left(\frac{q_{t,j}}{2\eta_{t,j}^2}\beta_{t,j}^2
-\psi(q_{t,j})\right)\right\}
\label{cost_aug}
\end{align}
where $\bm p_t=\{p_{t,i}\}_{i=1}^n$ and $\bm q_t=\{q_{t,j}\}_{j=1}^m$ are the collections of the auxiliary variables. As shown in Section~\ref{proof}, maximizing \eqref{cost_ro} with respect to $\bm x_t$ is equivalent to maximizing the augmented cost function (\ref{cost_aug}) in the enlarged parameter domain, i.e., $\{\bm p_t,\bm q_t, \bm x_t\}$. The augmented optimization problem can be solved in an alternating maximization manner. Specifically, we first maximize (\ref{cost_aug}) with respect to $\bm p_t$ and $\bm q_t$ while keeping $\bm x_t$ fixed. Then $\bm x_t$ is updated with the newest $\bm p_t$ and $\bm q_t$. This process is repeated until a convergence criterion is met.

For the first step of the $k$-th iteration after $\bm x_t^{k-1}$ is given and $\{\bm \alpha^{k-1}_{t},\ \bm \beta^{k-1}_{t}\}$ are computed via~\eqref{def1}\eqref{def2}, by Proposition~\ref{propo}, $\bm p_t^k$ and $\bm q_t^k$ are updated as follows,
\begin{align}
p_{t,i}^k&=-
\exp\left({-\frac{(\alpha_{t,i}^{k-1})^2}{2\sigma_{t,i}^2}}\right)\label{p_up}\\
q_{t,j}^k&=-
\exp\left({-\frac{(\beta_{t,j}^{k-1})^2}{2\eta_{t,j}^2}}\right)\label{q_up}
\end{align}
At the beginning of the iterative procedure, i.e., $k=1$, we set $\bm \alpha^{0}_{t}=\bm \beta^{0}_{t}=0$ as the initial condition.

For the second step of the $k$-th iteration, we fix $\{\bm p_t^k,\bm q_t^k\}$ and drop the items that do not depend on $\bm x_t$, and then (\ref{cost_aug}) can be simplified as
\begin{align}
\hat{\bm x}_{t|t}^k=\text{arg}\max_{\bm x_t}\Big(
\sum_{i=1}^n\frac{a_{t,i}p^k_{t,i}}{2\sigma_{t,i}^2}\alpha_{t,i}^2
+\sum_{j=1}^m\frac{b_{t,j}q^k_{t,j}}{2\eta_{t,j}^2}\beta_{t,j}^2
\Big)
\label{cost_last}
\end{align}

Before optimizing the above cost function, it is necessary to discuss the choices of the constants $a_{t,i}$ and $b_{t,j}$. In order to maintain the consistency of the Gaussian approximation filter, $a_{t,i}$ and $b_{t,j}$ should be properly selected so that (\ref{cost_last}) is equivalent to (\ref{cost:2}) when both process and measurement noises in the SSM are Gaussian. Specifically, when both noises follow the Gaussian distribution, the following constraints must be in place
\begin{align}
\frac{a_{t,i}p^k_{t,i}}{\sigma_{t,i}^2}&=-1\\
\frac{b_{t,j}q^k_{t,j}}{\eta_{t,j}^2}&=-1
\end{align}
In such a case, all bandwidths should be chosen as large as possible so that the correntropy behaves like the $\ell_2$ norm, and it follows from Proposition \ref{propo} that we can approximate $\{\bm p^k_{t,i}, \bm q^k_{t,j}\}$ by constant $-1$. Then the coefficients are obtained as follows
\begin{align}
{a_{t,i}}&={\sigma_{t,i}^2},\label{k1}\\
{b_{t,j}}&={\eta_{t,j}^2}
\label{k2}
\end{align}
Substituting \eqref{k1} and \eqref{k2} into \eqref{cost_last} leads to the updated cost function:
\begin{align}
\hat{\bm x}_{t|t}^k=\text{arg}\max_{\bm x_t}\left(
\sum_{i=1}^n\frac{p^k_{t,i}}{2}\alpha_{t,i}^2
+\sum_{j=1}^m\frac{q^k_{t,j}}{2}\beta_{t,j}^2\right)
\label{cost_last1}
\end{align}
which can equivalently expressed in a matrix format as
\begin{align}
\hat{\bm x}_{t|t}^k=\text{arg}\min_{\bm x_t}\left(
\frac{1}{2}\bm \alpha_{t}^T\bm \Psi_t\bm \alpha_{t}+\frac{1}{2}\bm \beta_{t}^T\bm \Phi_t\bm \beta_{t}\right)
\label{cost_last2}
\end{align}
where $\bm \Psi_t\in\mathbb{R}^{n\times n}$ and $\bm \Phi_t\in\mathbb{R}^{m\times m}$ are diagonal matrices given by
\begin{align}
\bm \Psi_t&=\text{diag}(-p_{t,1}^k,\cdots,-p_{t,n}^k)\label{ww1}\\
\bm \Phi_t&=\text{diag}(-q_{t,1}^k,\cdots,-q_{t,m}^k)\label{ww2}
\end{align}
Furthermore, substituting \eqref{def1} and \eqref{def2} into \eqref{cost_last2}, we obtain
\begin{align}
\hat{\bm x}_{t|t}^k=\text{arg}\min_{\bm x_t}\Big(\frac{1}{2}\|\bm x_t-\hat{\bm x}_{t|t-1}\|^2_{\bar{\bm P}_{t|t-1}^{-1}}+\frac{1}{2}\|\bm y_t-h(\bm x_t)\|^2_{\bar{\bm R}_{t}^{-1}}\Big)
\label{cost_last3}
\end{align}
where
\begin{align}
&\bar{\bm P}_{t|t-1}={\bm P}_{t|t-1}^{1/2}\bm \Psi_t^{-1}{\bm P}_{t|t-1}^{T/2}\label{P_upt}\\
&\bar{\bm R}_{t}={\bm R}_{t}^{1/2}\bm \Phi_t^{-1}{\bm R}_{t}^{T/2}\label{R_upt}
\end{align}

We notice that, expect for a different weighting matrix, \eqref{cost_last3} is similar to (\ref{cost2}), which can be efficiently solved in the Kalman filtering framework. Hence the second step of the $k$-th iteration can be implemented in the Kalman filtering framework with the modified weighting matrices. Once the $\hat{\bm x}_{t|t}^k$ is obtained, we can update the weighting matrices and then solve the problem to obtain $\hat{\bm x}_{t|t}^{k+1}$. The iterative procedure is repeated until a convergence criterion is met. The resulting robust cubature Kalman filter (RCKF) is summarized in Algorithm 1.

\begin{algorithm}[!t]
  \caption{Robust Cubature Kalman Filter Based on MCC} \label{alg:lrkf}
  \begin{algorithmic}[0]
    \State \textbf{Input:} $\bm y_{1:T}$, $\hat{\bm x}_{0|0}$, $\bm P_{0|0}$, $\bm Q_{1:T}$, $\bm R_{1:T}$.
    \State \textbf{Output:} $\bm \hat{\bm x}_{t|t}$ and $\bm P_{t|t}$ for $t=1:T$.
  \For{$t = 1,\cdots,T$}
  \State 1. Run the prediction step of CKF to calculate $\hat{\bm x}_{t|t-1}$
  \State \ and $\bm P_{t|t-1}$;
  \State 2. Initialize $\bm \Psi_t^{0}=\bm I_n,\ \bm \Phi_t^{0}=\bm I_m,\ k=0;$
  \Repeat
    \State 1. Update $\bar{\bm {P}}_{t|t-1}$ via \eqref{P_upt} and $\bar{\bm R}_t$ via \eqref{R_upt};
    \State 2. Update the innovation covariance $\bm P_{yy}$ and the
    \State  \quad\ filtering gain $\bm K_{t}$ via \eqref{G_upt} and \eqref{I_upt};
    \State 3. Calculate the filtering state $\hat{\bm x}_{t|t}^{k}$ via \eqref{s_upt};
    \State 4. Update $\bm\Phi_t$ and $\bm\Psi_t$ via \eqref{def1}, \eqref{def2},\eqref{p_up}, \eqref{q_up}, \eqref{ww1} and
    \State \quad \eqref{ww2};
    \State 5. $k=k+1$
  \Until a convergence criterion is met
  \State $\bm \hat{\bm x}_{t|t} = \hat{\bm x}_{t|t}^{k-1};\ \bm P_{t|t}=\bm P_{t|t-1}-\bm K_t\bm P_{yy}\bm K_t^T$.
  \EndFor
  \end{algorithmic}
\end{algorithm}

A closer examination of the proposed algorithm reveals that it is similar to the conventional Gaussian approximation Kalman filter, with the same prediction update step and iterative measurement update step. As proven by simulation results, the iteration number is relatively small in general (see Section~\ref{simu} for more details). Hence the proposed MCC based robust nonlinear Kalman filter has only a slightly higher computational complexity compared with the corresponding Gaussian approximation filer.

\subsection{Convergence}

\label{proof}

In this section, we provide a convergence analysis for the MCC based robust nonlinear Kalman filter. For brevity, we denote the original objective function~\eqref{cost_ro} and the transformed cost function~\eqref{cost_aug} by $\mathcal{Q}(\bm x_t)$ and $\mathcal{J}(\bm x_t,\bm p_t,\bm q_t)$, respectively. We first show that $\mathcal{Q}(\bm x_t)$ and $\mathcal{J}(\bm x_t,\bm p_t,\bm q_t)$ have the same optimal solution. We assume that $\bm x_t^+$ and $\{\bm x_t^*,\bm p_t^*,\bm q_t^*\}$ are the optimal solutions to $\mathcal{Q}(\bm x_t)$ and $\mathcal{J}(\bm x_t,\bm p_t,\bm q_t)$, respectively; in addition, $\{\bm p_t^+,\bm q_t^+\}$ maximize $\mathcal{J}(\bm x_t,\bm p_t,\bm q_t)$ when $\bm x_t$ equals $\bm x_t^+$. It is clear that
\begin{align}
\mathcal{Q}(\bm x_t^+)&\ge\mathcal{Q}(\bm x_t^*)\label{pro_1}\\
\mathcal{J}(\bm x_t^*,\bm p_t^*,\bm q_t^*)&\ge\mathcal{J}(\bm x_t^+,\bm p_t^+,\bm q_t^+)
\label{pro_2}
\end{align}
According to Proposition 1, the following equalities hold
\begin{align}
\mathcal{Q}(\bm x_t^+)&=\mathcal{J}(\bm x_t^+,\bm p_t^+,\bm q_t^+)\label{pro_3}\\
\mathcal{Q}(\bm x_t^*)&=\mathcal{J}(\bm x_t^*,\bm p_t^*,\bm q_t^*)
\label{pro_4}
\end{align}
Combing~\eqref{pro_1}-\eqref{pro_4}, we obtain that $\mathcal{Q}(\bm x_t^+)=\mathcal{Q}(\bm x_t^*)$, which illustrates that the original optimization problem (\ref{cost_ro}) and the transformed one (\ref{cost_aug}) have the same optimal solution.

Next, we prove that the sequence $\{\bm x_{t|t}^k,\bm p_t^k,\bm q_t^k\}$ generated by Algorithm 1 converges, i.e.,
\begin{align}
\lim\limits_{k\rightarrow\infty}\left(\mathcal{J}(\bm x_{t|t}^{k+1},\bm p_t^{k+1},\bm q_t^{k+1})-\mathcal{J}(\bm x_{t|t}^k,\bm p_t^k,\bm q_t^k)\right)=0
\label{conn}
\end{align}
According to Algorithm 1, the following equations hold
\begin{align}
\mathcal{J}(\bm x_{t|t}^k,\bm p_t^{k+1}\bm q_t^{k+1})&\ge\mathcal{J}(\bm x_{t|t}^k,\bm p_t^k,\bm q_t^k)\label{11}\\
\mathcal{J}(\bm x_{t|t}^{k+1},\bm p_t^{k+1}\bm q_t^{k+1})&\ge\mathcal{J}(\bm x_{t|t}^k,\bm p_t^{k+1}\bm q_t^{k+1})\label{22}
\end{align}
Combining (\ref{11}) and (\ref{22}), we conclude that the sequence $\mathcal{J}(\bm x_{t|t}^k,\bm p_t^k,\bm q_t^k)$ produced by Algorithm 1 is non-decreasing. Furthermore, the objective function (\ref{cost_aug}) is bounded above. According to the monotone convergence theorem~\cite{James:2006}, we can draw the conclusion that (\ref{conn}) holds.

To sum up, our proposed robust filtering algorithm yields the same solution of the original problem through the half-quadratic technique and  its iteratively generated sequence converges. The proof of convergence is now complete.

\section{\!Robust Nonlinear Kalman Smoother with MCC}
\label{smcc}
In this section, we derive the MCC based robust Kalman smoother along with the similar clue in Section~\ref{Sec:fmcc}. In the Bayesian smoothing paradigm, the posterior density of the states $p(\bm x_{0:T}|\bm y_{1:T})$ is given by
\begin{align}
p(\bm x_{0:T}|\bm y_{1:T})\propto p(\bm x_0)\prod_{t=1}^Tp(\bm x_{t}|\bm x_{t-1})\prod_{t=1}^Tp(\bm y_t|\bm x_t)
\end{align}
The MAP estimate of the smoothed states is obtained by minimizing the negative log posterior density $p(\bm x_{0:T}|\bm y_{1:T})$, i.e.,
\begin{align}
\hat{\bm x}_{0:T|T}=\arg\min_{{\bm x}_{0:T}}\Big(-\log\left(p(\bm x_0)\right)-\sum_{t=1}^T\log\left(p(\bm x_{t}|\bm x_{t-1})\right)-\sum_{t=1}^T\log\left(p(\bm y_t|\bm x_t)\right)\Big)
\end{align}
Under the Gaussian assumption and dropping the terms that do not relate on ${\bm x}_{0:T}$, $\hat{\bm x}_{0:T|T}$ can be obtained by
\begin{align}
\hat{\bm x}_{0:T|T}=\arg\min_{{\bm x}_{0:T}}\Big(
\frac{1}{2}\sum_{t=1}^T\|\bm x_t-f(\bm x_{t-1})\|^2_{{\bm Q}_{t-1}^{-1}}+\frac{1}{2}\sum_{t=1}^T\|\bm y_t-h(\bm x_t)\|^2_{{\bm R}_{t}^{-1}}+\frac{1}{2}\|\bm x_0-\hat{\bm x}_{0|0}\|_{{\bm P_{0|0}^{-1}}}^2\Big)
\label{cost:sm}
\end{align}
Similarly, (\ref{cost:sm}) can be written in terms of fitting errors as
\begin{align}
\hat{\bm x}_{0:T|T}&=\arg\min_{{\bm x}_{0:T}}\left(
\frac{1}{2}\sum_{t=0}^T\bm\alpha_t^T\bm\alpha_t
+\frac{1}{2}\sum_{t=1}^T\bm\beta_t^T\bm\beta_t\right)\notag\\
&=\text{arg}\min_{{\bm x}_{0:T}}\left(\frac{1}{2}\sum_{t=0}^T\sum_{i=1}^n\alpha_{t,i}^2
+\frac{1}{2}\sum_{t=1}^T\sum_{j=1}^m\beta_{t,j}^2\right)
\label{cost:sm_e}
\end{align}
where $\bm \beta_t$ has a same definition in~\eqref{def2}, $\bm \alpha_t$ is redefined as
\begin{align}
\bm \alpha_t=\left\{
\begin{array}{ll}
\bm P_{0|0}^{-1/2}\left(\bm x_0-\hat{\bm x}_{0|0}\right),&t=0\\
\bm Q_{t-1}^{-1/2}\left(\bm x_t-h(\bm x_{t-1})\right),&t\ne 0
\end{array}\right.
\label{def0}
\end{align}

Akin to the robust nonlinear Kalman filter, we replace the quadratic loss with the weighted instantaneous correntropy for each component of the normalized error, resulting in the following objective function for the nonlinear robust smoother
\begin{align}
\hat{\bm x}_{0:T|T}=\text{arg}\max_{{\bm x}_{0:T}}\Big( \sum_{t=0}^T\sum_{i=1}^n &a_{t,i}\kappa_{\sigma_{t,i}}(\alpha_{t,i})+\sum_{t=1}^T\sum_{j=1}^m b_{t,j}\kappa_{\eta_{t,j}}(\beta_{t,j})\Big)
\label{cost:sr_cor}
\end{align}
where $\{a_{t,i},b_{t,j}\}$ are parameters that should be selected to maintain the consistency of the Gaussian approximation smoother when the SSM does not involve the impulsive noises, similarly to the filtering case discussed in Section~\ref{par_dis}. According to Proposition 1, \eqref{cost:sr_cor} can be transformed as
\begin{align}
\!\!\hat{\bm x}_{0:T|T}=\text{arg}\max_{{\bm x},{\bm p},{\bm q}}&\Bigg( \sum_{t=0}^T\sum_{i=1}^n a_{t,i}\Big(\frac{p_{t,i}}{2\sigma_{t,i}^2}\alpha_{t,i}^2-\psi(p_{t,i})\Big)+\sum_{t=1}^T\sum_{j=1}^m b_{t,i}\Big(\frac{q_{t,i}}{2\eta_{t,i}^2}\beta_{t,i}^2-\psi(q_{t,i})\Big)\Bigg)
\label{cost:sr_cor1}
\end{align}
where $\bm x=\{\bm x_t\}_{t=0}^T$, $\bm p=\{\bm p_t\}_{t=0}^T$ and $\bm q=\{\bm q_t\}_{t=1}^T$ are the collections of all variables to be optimized.

Again, the augmented objective function can be solved in an iterative manner via the half-quadratic method. First of all, fixing the states and optimizing~\eqref{cost:sr_cor1} with respect to $\bm p$ and $\bm q$ according to Proposition 1, we obtain the same updating equations for $\{p_{t,i},\ q_{t,j}\}$ as shown in~\eqref{p_up}\eqref{q_up}. Next, keeping $\bm p$ and $\bm q$ fixed and dropping the items that do not relate to the state, one can update the smoothed state $\hat{\bm x}_{0:T|T}$ though
\begin{align}
\hat{\bm x}_{0:T|T}=\text{arg}\max_{{\bm x_{0:T}}}& \Big(\sum_{t=0}^T\sum_{i=1}^n \frac{p_{t,i}a_{t,i}}{2\sigma_{t,i}^2}\alpha_{t,i}^2+\sum_{t=1}^T\sum_{j=1}^m \frac{q_{t,i}b_{t,i}}{2\eta_{t,i}^2}\beta_{t,i}^2\Big)
\label{cost:sr_cor2}
\end{align}
Similarly, to maintain the consistency of the Gaussian approximation smoother when the SSM dose not involve any impulsive noise, the coefficients should be selected by~\eqref{k1} and \eqref{k2}. Substituting \eqref{k1} and \eqref{k2} into (\ref{cost:sr_cor2}), the objective function is updated as
\begin{align}
\hat{\bm x}_{0:T|T}=\text{arg}\max_{{\bm x_{0:T}}}\Big( \sum_{t=0}^T\sum_{i=1}^n&\frac{1}{2} p_{t,i}\alpha_{t,i}^2+\sum_{t=1}^T\sum_{j=1}^m \frac{1}{2}q_{t,i}\beta_{t,i}^2\Big)
\label{cost:sr_cor3}
\end{align}
Writing in matrix format, we get
\begin{align}
\hat{\bm x}_{0:T|T}=\text{arg}\min_{\bm x_{0:T}}\left(
\frac{1}{2}\sum_{t=0}^T\bm \alpha_{t}^T\bm \Psi_t\bm \alpha_{t}+\frac{1}{2}\sum_{t=1}^T\bm \beta_{t}^T\bm \Phi_t\bm \beta_{t}
\right)
\label{obj_last2}
\end{align}
where $\bm \Psi_t\in\mathbb{R}^{n\times n}$ and $\bm \Phi_t\in\mathbb{R}^{m\times m}$ are diagonal matrixes, given by \eqref{ww1} and \eqref{ww2}, respectively. Furthermore, substituting the expression of $\bm \alpha_{t}$ and $\bm \beta_{t}$, namely \eqref{def0} and \eqref{def2}, into (\ref{obj_last2}), we obtain
\begin{align}
\hat{\bm x}_{0:T|T}=\text{arg}\min_{\bm x_{0:T}}\Big(\frac{1}{2}\sum_{t=1}^T\|\bm x_t-f({\bm x}_{t-1})\|^2_{\bar{\bm Q}_{t-1}^{-1}}+\frac{1}{2}\sum_{t=1}^T\|\bm y_t-h(\bm x_t)\|^2_{\bar{\bm R}_{t}^{-1}}+\frac{1}{2}\|\bm x_0-\hat{\bm x}_{0|0}\|_{\bar{\bm P}_{0|0}^{-1}}^2\Big)
\label{obj_last3}
\end{align}
where
\begin{align}
&\bar{\bm Q}_{t-1}={\bm Q}_{t-1}^{1/2}\bm \Psi_t^{-1}{\bm Q}_{t-1}^{T/2}\label{w1}\\
&\bar{\bm R}_{t}={\bm R}_{t}^{1/2}\bm \Phi_t^{-1}{\bm R}_{t}^{T/2}\\
&\bar{\bm P}_{0|0}={\bm P}_{0|0}^{1/2}\bm \Psi_0^{-1}{\bm P}_{0|0}^{T/2}\label{w2}
\end{align}

The objective function (\ref{obj_last3}) can be solved by CKS, which provides updates of the state $\hat{\bm x}_{0:T|T}$. The iterative procedure is repeated till some convergence criterion is met. The detailed robust cubature Kalman smoother (RCKS) is summarized in Algorithm 2.

\begin{algorithm}[thbp]
  \caption{MCC Based Robust Cubature Kalman Smoother} \label{alg:lrkf}
  \begin{algorithmic}[0]
    \State \textbf{Input:} $\bm y_{1:T}$, $\hat{\bm x}_0$, $\bm P_0$, $\bm Q_{1:T}$, $\bm R_{1:T}$.
    \State \textbf{Output:} $\bm \hat{\bm x}_{t|T}$ and $\bm P_{t|T}$ for $t=1:T$.
    \State \textbf{Initialized:} $k=0$, $\bm \Psi_t=\bm I_n$, $\bm \Phi_t=\bm I_m$.
  \Repeat
  \State 1. Update $\bar{\bm Q}_{t-1},\bar{\bm R}_{t}\ \text{and }\bar{\bm P}_0$ via \eqref{w1}-\eqref{w2};
  \State 2. Update $\hat{\bm x}_{0:T|T}^k\ \text{and }{\bm P}_{0:T|T}^k$ via CKS;
  \State 3. Update $\bm\Phi_t$ and $\bm\Psi_t$ via \eqref{def0}, \eqref{def2}, \eqref{p_up}, \eqref{q_up}, \eqref{ww1} and
  \State  \quad \eqref{ww2};
  \State 4. $k=k+1$;
  \Until a convergence criterion is met;
  \State $\bm \hat{\bm x}_{t|T}=\bm \hat{\bm x}_{t|T}^{(k-1)}$ and $\bm P_{t|T}=\bm P_{t|T}^{(k-1)}$;
  \end{algorithmic}
\end{algorithm}

It can be seen that the proposed robust nonlinear Kalman smoother actually contains two steps in each iteration, i.e., updating the noise covariance and running the conventional Gaussian approximation smoother. One can follow similar steps used in Section~\ref{proof} and show that the proposed nonlinear robust Kalman smoother converges to a local minimum of the cost function.

\section{Numerical Simulations}
\label{simu}
Numerical examples are presented in this section to \mbox{illustrate} the performance of the proposed RCKF and RCKS in the presence of heavy-tailed noises. We also compare the state estimates of the proposed algorithms with those of the conventional CKF~\cite{arasaratnam:2009}, conventional CKS~\cite{arasaratnam:2011}, linear regression and MCC based robust Kalman filter (LRKF)~\cite{liu:2017}, nonlinear regression based Huber Kalman filter (HRKF)~\cite{karlgaard:2015} and variational Bayesian based student's cubature Kalman smoother (TCKS)~\cite{huang:2016_2}. The LRKF is originally proposed in the unscented Kalman filter (UKF) framework and here for consistency with the other compared algorithms, we set parameters $(\alpha,\beta,\kappa)=(1,0,0)$ in LRKF, which makes it functionally equivalent to a CKF. Other design parameters in the aforementioned algorithms are set as recommended in their literatures. In the simulation, we take the following convergence criterion, $\|\hat{\bm x}_{t|t}^{k}-\hat{\bm x}_{t|t}^{k-1}\|/\|\hat{\bm x}_{t|t}^{k-1}\|\le 10^{-6}$ in filtering and $\text{max}(\|\hat{\bm x}_{t|T}^{k}-\hat{\bm x}_{t|T}^{k-1}\|/\|\hat{\bm x}_{t|T}^{k-1}\|)\le 10^{-6}$ for $t=1,\cdots,T$ in smoothing.

\subsection{Van der Pol Oscillator}

The Van der Pol oscillator (VPO) model~\cite{Kandepu:2008} is usually used as a benchmark for testing the performance of nonlinear filters and smoothers. The continuous-time nonlinear dynamics of the VPO is governed by the following differential equations
\begin{align}
\dot {x}_1&=x_2\\
\dot {x}_2 &= \mu\left(1-x_1^2\right)x_2-x_1
\end{align}
where $\mu$ is a scalar parameter indicating the nonlinearity. Discretization of the VPO yields
\begin{align}
\!\!\bm x_{t+1} =
\left(
\begin{array}{>{\displaystyle}l}
\!\!x_{1,t}+\int_{t}^{t+\delta}x_2dt\!\!\!\!\\
\!\!x_{2,t}+\int_{t}^{t+\delta}\left(\mu\left(1-x_1^2\right)x_2-x_1\right)dt\!\!\!\!
\end{array}
\right)+\bm\omega_t
\label{VPO_sys}
\end{align}
where $\bm x_t=[x_{1,t},x_{2,t}]^T$ is the state, $\bm \omega_t$ is the process noise to model the discretization error, and $\delta$ is the sampling interval. The integral terms in \eqref{VPO_sys} are numerically calculated by the fourth-order Runge-Kutta scheme with a time step of the sampling interval. Furthermore, we assume that the state is measured by the following noise disturbed nonlinear procedure
\begin{align}
y_t = (x_{1,t}-1)^2+1+v_{t}
\end{align}
Both noises are generated from the mixed-Gaussian model~\cite{Huang:2016}:
\begin{align}
\bm\omega_{t-1}&\thicksim(1-p_1)\mathcal{N}(0,\bm Q_{t-1})+p_1\mathcal{N}(0,\phi_1\bm Q_{t-1})\\
v_{t}&\thicksim(1-p_2)\mathcal{N}(0,R_t)+p_2\mathcal{N}(0,\phi_2R_t)
\end{align}
where $p_1$ and $p_2$ are the contaminating parameters, $\phi_1$ and $\phi_2$ are the scaling factors indicating the strength of the contaminating noises, and $\bm Q_{t-1}$ and $R_t$ are the covariance matrices of the nominal process noise and measurement noise, respectively.

In the simulation, we select $\mu=1$ and total samples $T=120$ with the sampling interval $\delta=0.1$s. The initial state $\hat{\bm x}_{0|0}$ follows $\mathcal{N}(\bm x_0,0.01\bm I_2)$ with $\bm x_0=[0,-0.5]^T$. The covariance of the noises are set as $\bm Q_{t-1}=0.01\bm I_2$ and $R_t=1$, respectively. The covariance of the contaminating part in each noise is controlled by $\phi_1=10$ and $\phi_2=50$. Specifically, three scenarios are considered: 1) all noises are Gaussian, i.e., $p_1=p_2=0$; 2) the process noise is a Gaussian noise while the measurement noise is a heavy-tailed noise, i.e., $p_1=0$ and $p_2=0.2$; 3) both process and measurement noises are heavy-tailed noises, i.e., $p_1=p_2=0.2$.

For fair comparison, we implement $L=1000$ Monte Carol runs in each scenario. We use the implementation time (IT) and time-averaged root mean square error (TRMSE) as performance metrics. The TRMSE is defined as
\begin{align}
\text{TRMSE} = \frac{1}{T}\sum_{t=1}^T\sqrt{\frac{1}{L}\sum_{i=1}^L\left(x_t^i-\hat{x}_t^i\right)^2}
\label{trmse}
\end{align}
where $x_{t}^i$ and $\hat{x}_{t}^i$, respectively, denote the true and estimated state component at time $t$ in the $i$-th Monte Carlo run.

\begin{figure}[!h]
  \centering
  \includegraphics[width = 1.0\columnwidth]{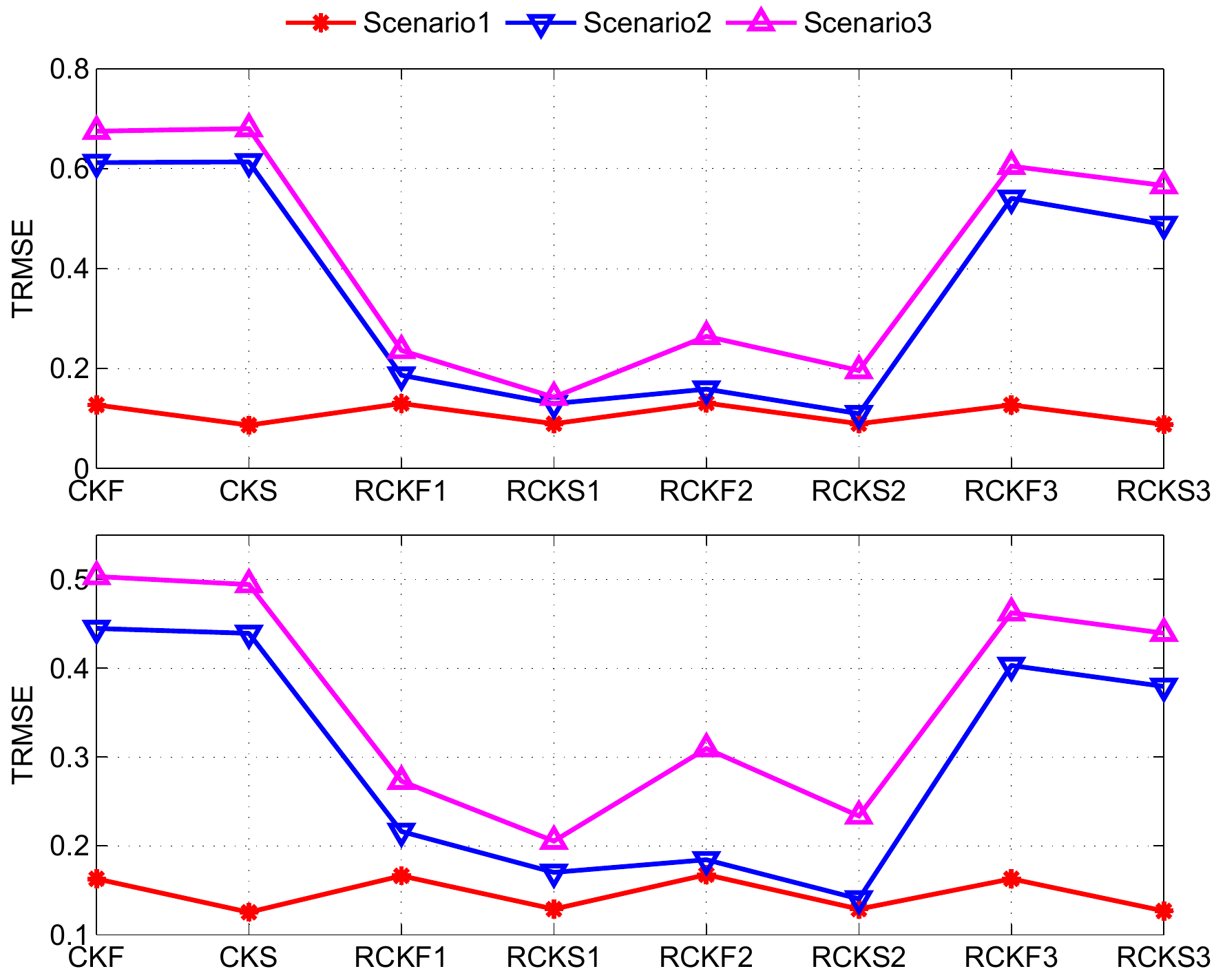}
  \caption{The TRMSEs of different methods for $x_1$ (top) and $x_2$ (bottom).}
  \label{fig_trmse1}
\end{figure}

\begin{figure}[!h]
  \centering
  \includegraphics[width = 1.0\columnwidth]{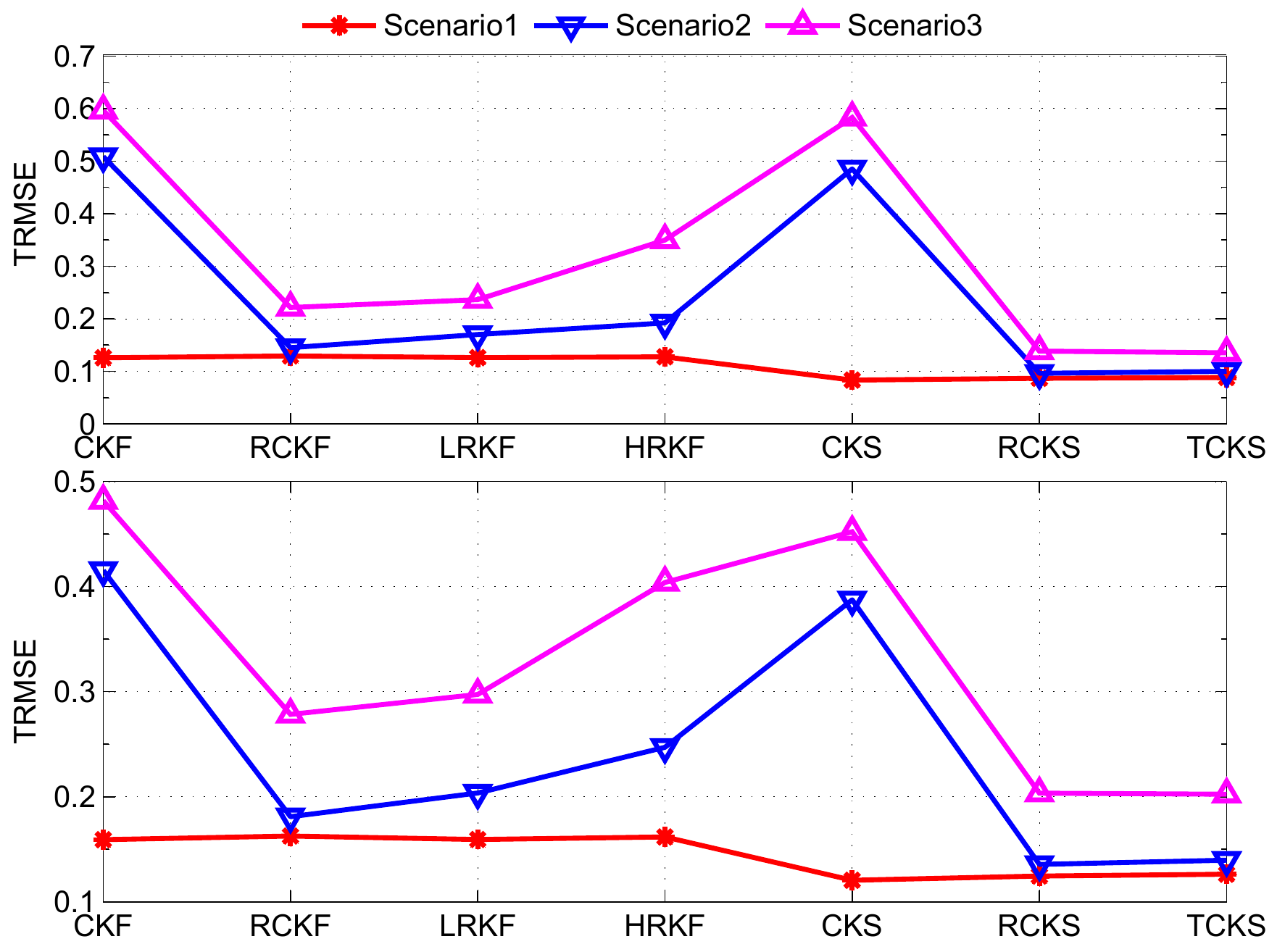}\label{fig_trmse2}
  \caption{The TRMSEs of the different methods for $x_1$ (top) and $x_2$ (bottom).}
  \label{fig:2}
\end{figure}

The kernel size is the only design parameter in the proposed algorithms and it has a significant impact on the performance of the RCKF and RCKS. Therefore, we studied the performance of the proposed algorithms with three different kernel sizes, i.e., $\sigma_t=\eta_t=2$; $\sigma_t=20,\ \eta_t=2$ and $\sigma_t=\eta_t=20$, and the results are labeled as RCKF1/RCKS1, RCKF2/RCKS2 and RCKF3/RCKS3, respectively.  The conventional CKF and CKS are also implemented as a performance benchmark.

Fig.~\ref{fig_trmse1} shows the TRMSEs while the IT is collected in Table~\ref{Table:1}. It can be seen, as expected, the smoothers have higher accuracies at the cost of more computational time than their filtering counterparts due to the use of backward procedure. The IT of the robust estimators with different kernel sizes in three scenarios are similar, which implies the IT of the proposed algorithms is insensitive to the kernel size.

\begin{table}[!h]
\centering
\caption{The implementation time of different methods}
\label{Table:1}
\begin{tabular}[width = 0.85\columnwidth]{@{\extracolsep{\fill}}c|ccc|c|ccc}
\hline
\multirow{2}{*}{Method} & \multicolumn{3}{c|}{IT/s} & \multirow{2}{*}{Method} & \multicolumn{3}{c}{IT/s} \\
\cline{2-4}\cline{6-8}
                        & S1     & S2    & S3    &                         & S1     & S2    & S3    \\
                        \hline
CKF                     &37.1& 36.8  & 36.6   & RCKF1                   & 142.6     & 140.9     &130.8  \\
CKS                     & 44.5      & 43.8     & 42.1     & RCKS1                   & 643.0      & 590.7     & 533.3     \\
RCKF2                   & 136.7      & 133.0     & 122.4     & RCKF3                   & 127.8      & 130.8     & 122.0     \\
RCKS2                   & 648.5      & 588.6     & 532.4     & RCKS3                   & 642.8      & 589.8     & 531.9\\
\hline
\end{tabular}
\footnotetext{a ha}
\end{table}

In the scenario 1, the robust algorithms are similar to the conventional ones, which suggest that the robust algorithms are consistent with the conventional ones. When heavy-tailed noises are involved, taking their characteristics into account in general pays off in terms of accuracy, while the computational time increase because of their iterative procedure. It is observed that CKF3/CKS3, which employ a larger kernel size, perform the worst among the robust algorithms. CKF1/CKS1 and CKF2/CKS2 are slightly different in their perfermance, which is caused by their different kernel sizes.

\begin{table}[!h]
\centering
\caption{The implementation time of different methods}
\label{Table:2}
\begin{tabular}[width = 0.85\columnwidth]{@{\extracolsep{\fill}}c|ccc|c|ccc}
\hline
\multirow{2}{*}{Method} & \multicolumn{3}{c|}{IT/s} & \multirow{2}{*}{Method} & \multicolumn{3}{c}{IT/s} \\
\cline{2-4}\cline{6-8}
                        & S1     & S2    & S3    &                         & S1     & S2    & S3    \\
                        \hline
CKF                     & 37.1& 36.8  & 36.6 &    CKS                   & 44.5      & 43.8     & 42.1 \\
RCKF                     & 136.7      & 133.0     & 132.4     & RCKS                   & 648.5      & 588.6     & 532.4    \\
LRKF                   & 95.1      & 98.3     & 90.4     &  TCKS                   & 1097.2      & 964.4     & 941.1     \\
HRKF                   & 66.1      & 68.1     & 63.9     &              &       &   &\\
\hline
\end{tabular}
\label{T2}
\footnotetext{a ha}
\end{table}

\begin{figure*}[!t]
\centering
\subfloat[]{
\includegraphics[width=0.5\textwidth]{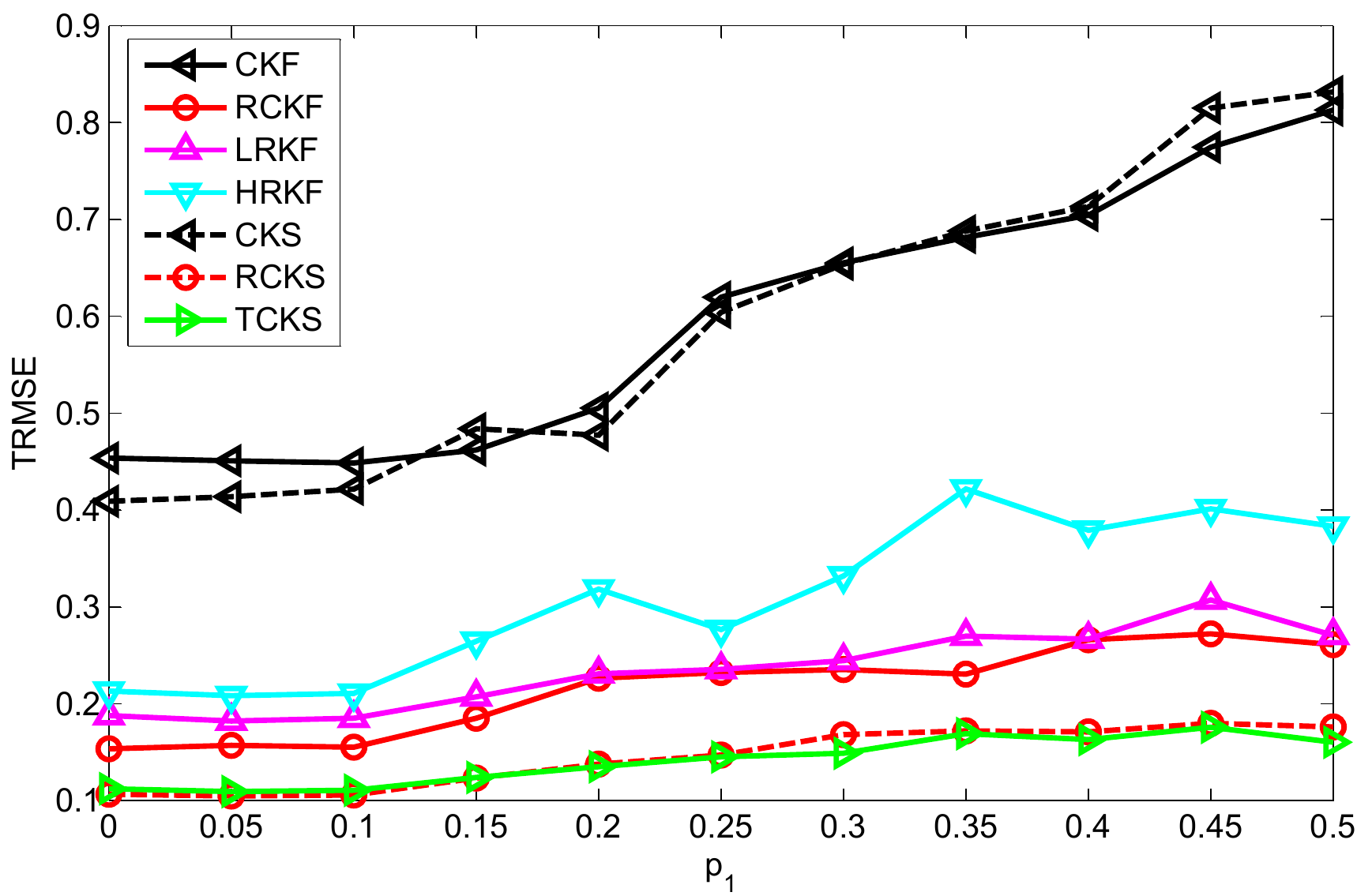}}
\subfloat[]{
\includegraphics[width=0.5\textwidth]{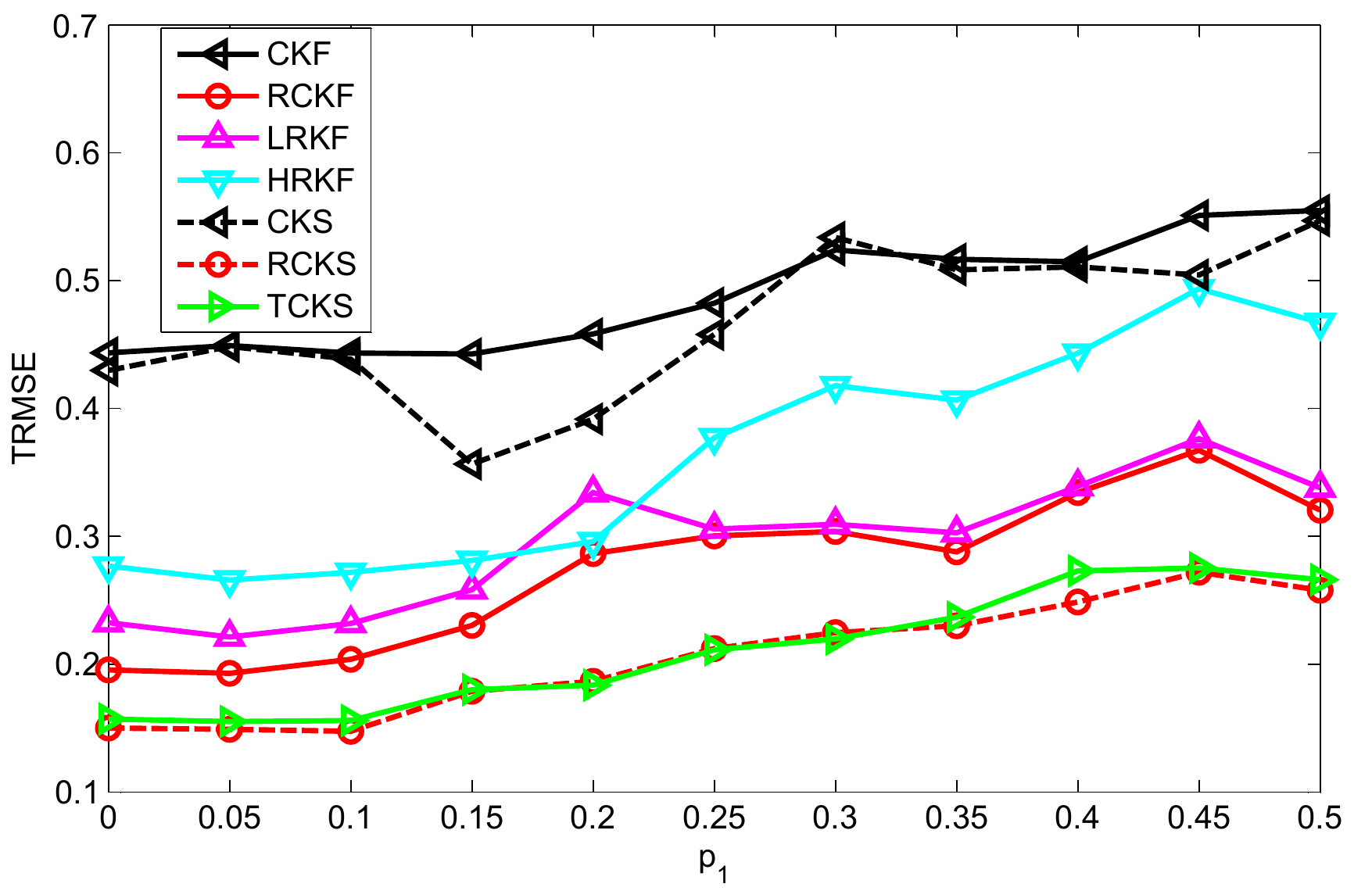}}
\\
\subfloat[]{
\includegraphics[width=0.5\textwidth]{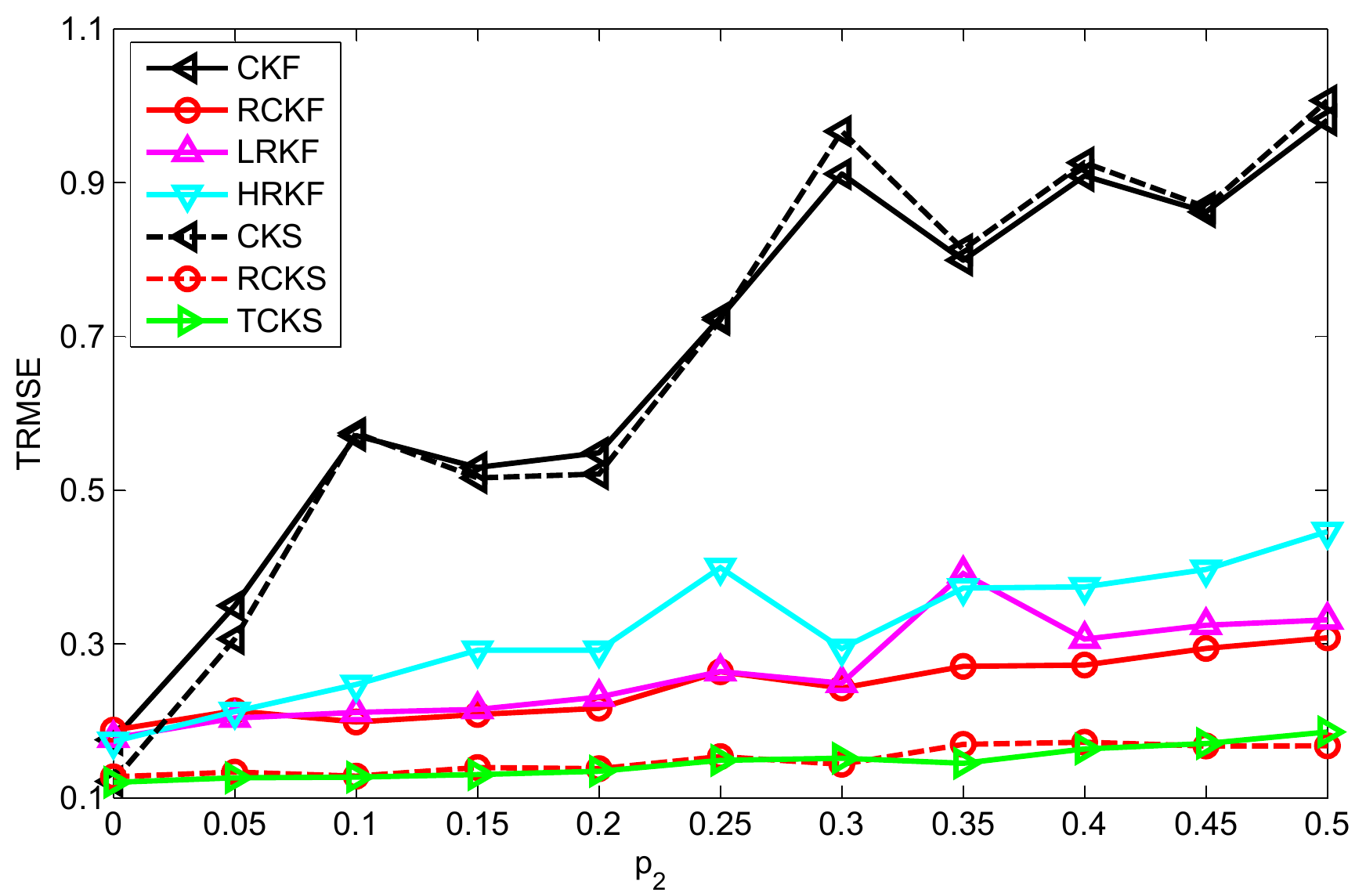}}
\subfloat[]{
\includegraphics[width=0.5\textwidth]{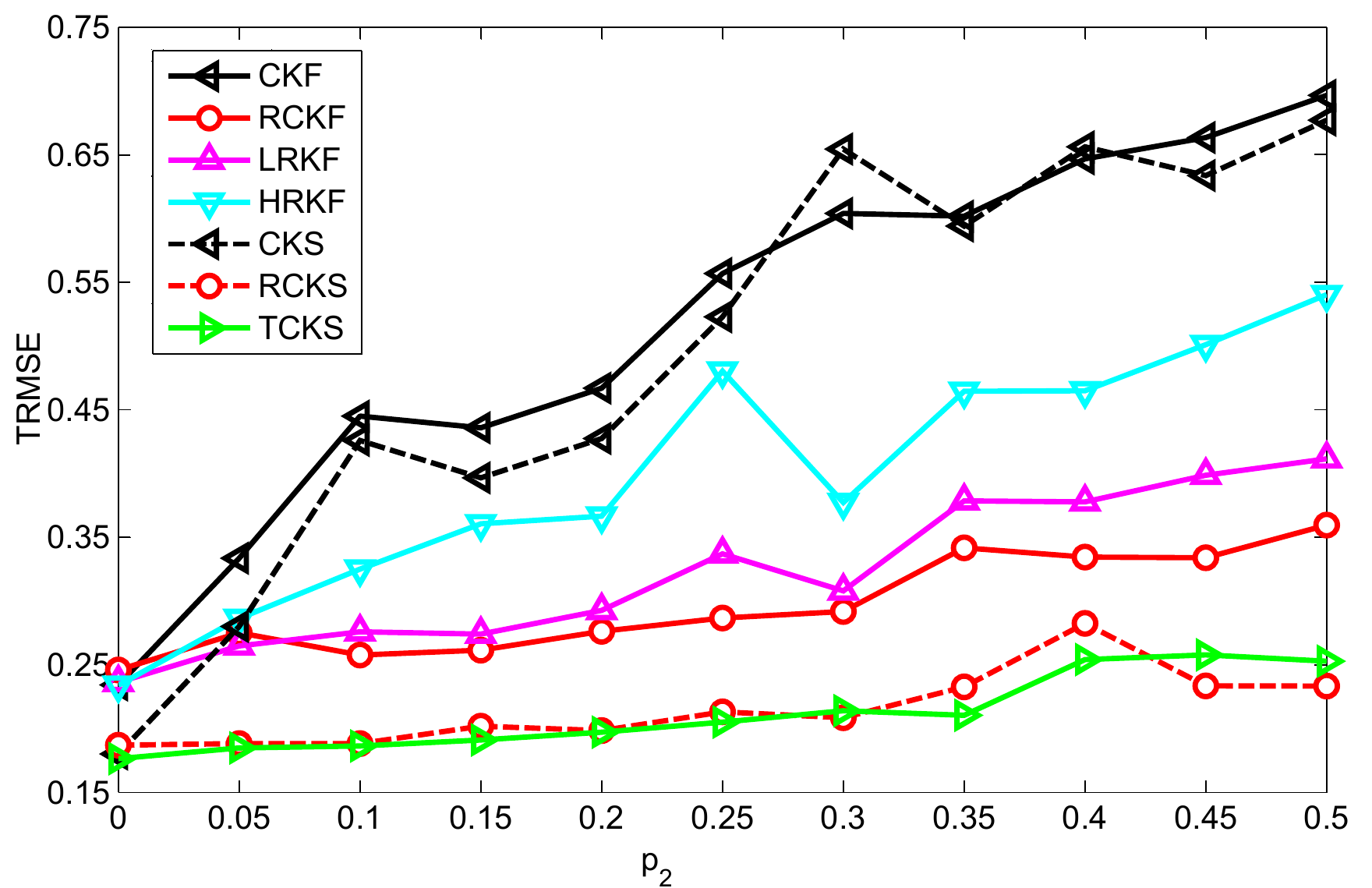}}
\caption{TRMSE of different estimators via the contamination ratio. Top: varying $p_1$ with fixed $p_2=0.2$, (a) TRMSE of the first component; (b) TRMSE of the second component. Bottom: varying $p_2$ with fixed $p_1=0.2$, (c) TRMSE of the first component; (d) TRMSE of the second component.}
\label{f3}
\end{figure*}

Next we present comparison results for different algorithms. For fairness, we set all kernel sizes to $2$ in LRKF, RCKF and RCKS. Fig.~\ref{fig:2} shows the TRMSEs of the different algorithms in the three scenarios and the IT of the different algorithms are given in Table~\ref{T2}. It is clear that all robust algorithms have a similar performance as that of the conventional estimators in the scenario when all noises are Gaussian. However, in the presence of the heavy-tailed noises, the robust estimators have a significant improvement over the conventional algorithms, which does not account for the presence of outliers. It is observed that the proposed RCKF outperforms HRKF. This is mainly because the MCC is instinctively more suitable for heavy-tailed noise than Huber's function. It is also seen that RCKF has a slight gain than LRKF. This is because, while both algorithms are derived from MCC, LRKF is based on a linear regression which introduces the linearzation error. Although RCKF is better than LRKF and HRKF in accuracy, the former has a higher IT than the latter, by $30\%$ and $95\%$, respectively. On the other hand, all three smoothers have smaller TRMSEs than their filtering counterparts. The proposed RCKS has a similar performance to that of TCKS and both outperform the traditional CKS. TCKS, however, needs about $40\%$ more IT than the proposed RCKS.

The TRMSE versus the increase of the contamination ratio is also studied, as shown in Fig.~\ref{f3}. The TRMSEs of the two traditional algorithms rise considerably, while the inverse is true for all robust algorithms except HRKF. The slightly increasing TRMSEs of the robust algorithms suggests that these algorithms are relatively insensitive to the level of the heavy-tailed noise. The TRMSEs of HRKF have a slight increase when the contamination ratio is relatively small and then the increase is more notable when the contamination ratio becomes larger, confirming that Huber's function based robust estimators performs well only at relativly low levels of contamination.

\subsection{Agile Target Tracking}

Here we consider the problem of tracking an agile target which is observed by an active radar located at the original point. The outliers may occur in both the process and measurement procedure due to the rapid motion of the target or disturbances experienced by the radar. The SSM is formulated as~\cite{li2003survey, bar2004estimation, arasaratnam2007discrete,huang:2016_2}
\begin{align}
\bm x_t &= \left[
\begin{array}{cc}
\bm I_2&\delta\bm I_2\\\bm 0 & \bm I_2
\end{array}
\right]\bm x_{t-1}+\bm\omega_{t-1} \\
\bm y_t &= \left[
\begin{array}{c}
\sqrt{a_t^2+b_t^2}\\ \text{atan2}(b_t,a_t)
\end{array}\right]
+\bm v_t
\end{align}
where $\bm x_t =[a_t,b_t,\dot{a}_t,\dot{b}_t]^T$ is the state; $a_t$, $b_t$, $\dot{a}_t$ and $\dot{b}_t$ denote the positions and corresponding velocities in Cartesian coordinates; $\delta=0.5$s is the sampling interval and $\text{atan2}$ is the four-quadrant inverse tangent function. In the simulation, the initial state $\hat{\bm x}_{0|0}$ is randomly chosen from $\mathcal{N}(\bm x_0,\bm P_0)$ with the true initial state $\bm x_0=[-10000;10000;30;-40]^T$ and corresponding error covariance $\bm P_0=100\bm I_4$. Outlier-corrupted process and measurement noises are generated from the following Gaussian mixture model:
\begin{align}
\bm\omega_{t-1}&\sim 0.8\mathcal{N}(0,\bm Q_{t-1})+0.2\mathcal{N}(0,10\bm Q_{t-1})\\
\bm v_t&\sim 0.8\mathcal{N}(0,\bm R_t)+0.2\mathcal{N}(0,50\bm R_t)
\end{align}
where the nominal covariance matrices for the process and measurement noise, i.e. $\bm Q$ and $\bm R$, are given as
\begin{align}
\!\!\!
\bm Q_{t-1}=\left[
\begin{array}{cc}
\frac{\delta^3}{3}\bm{I}_2&\!\frac{\delta^2}{2}\bm{I}_2\\
\frac{\delta^2}{2}\bm{I}_2&\!\delta\bm{I}_2
\end{array}\right],\
\bm R_t=\left[
\begin{array}{cc}
100\text{m}^2&\!\!0\\
0&\!\!16\text{mrad}^2
\end{array}\right]
\end{align}
$L=1000$ independent Monte-Carlo runs are implemented and in each run $T=200$ noisy measurements are collected. The root mean square error (RMSE) of position and velocity are performed as performance metrics, which are defined as
\begin{align}
\text{RMSE}_{\text{pos}}(t)&=\sqrt{\frac{1}{L}\sum_{i=1}^L\left(( a_{t}^i-\hat{ a}_{t}^i)^2+( b_{t}^i-\hat{ b}_{t}^i)^2\right)}
\end{align}
where $a_{t}^i$ and $\hat{ a}_{t}^i$ have similar meanings in~\eqref{trmse}. Akin to position, the RMSE of velocity can also be formulated.

\begin{figure}[]
\centering
\subfloat[]{
\includegraphics[width=0.45\textwidth]{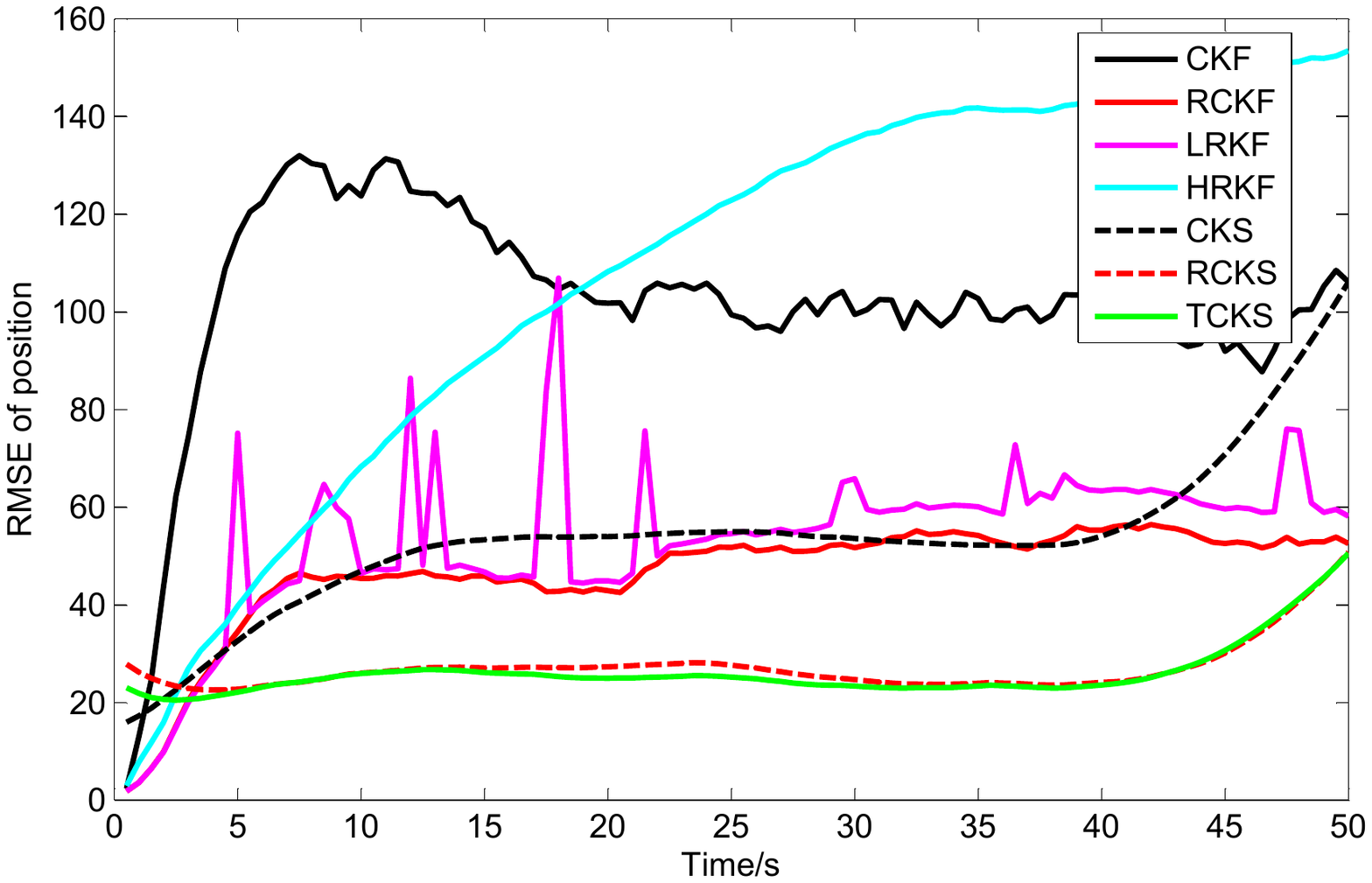}}\\
\subfloat[]{
\includegraphics[width=0.45\textwidth]{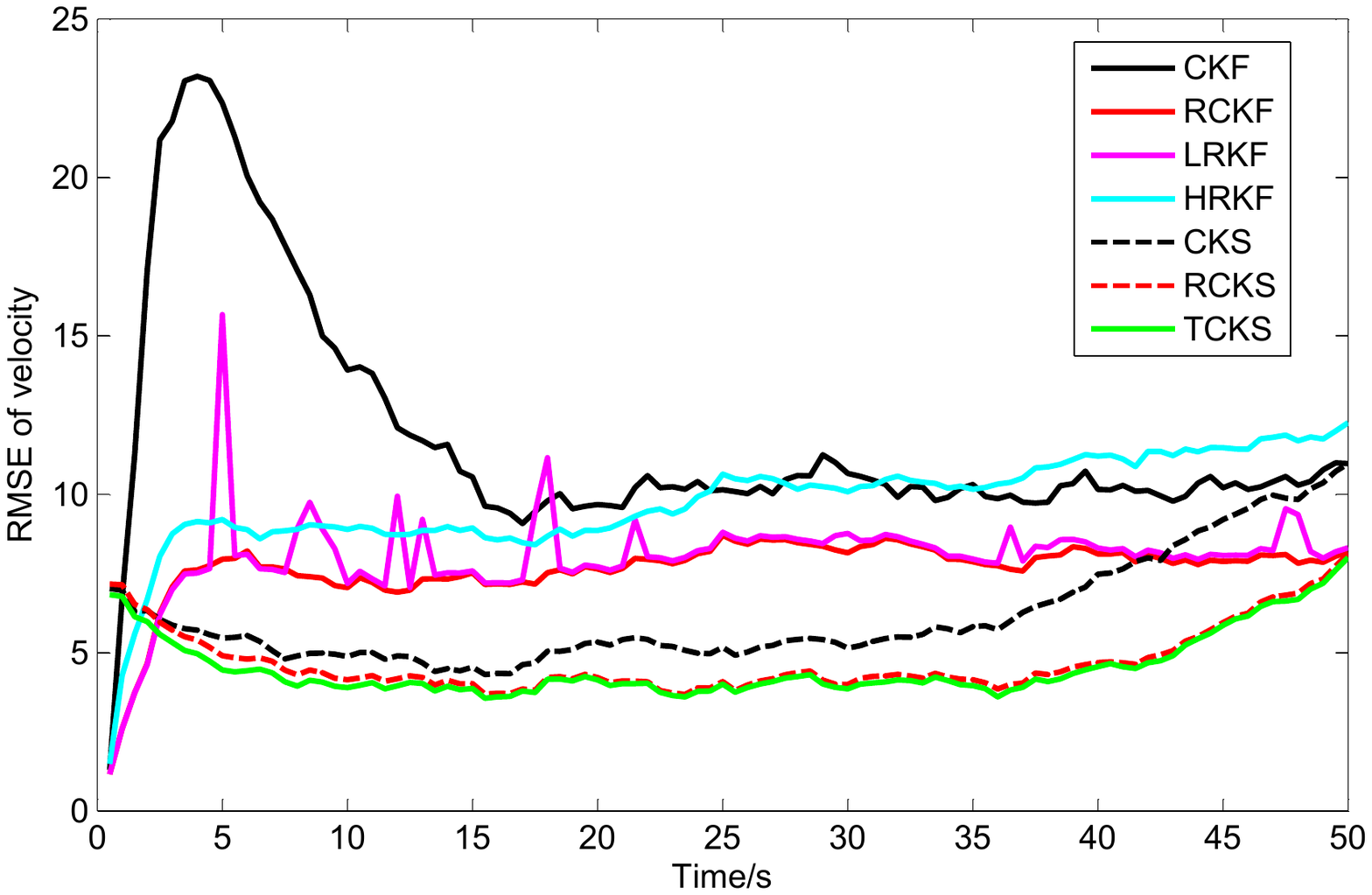}}
\caption{The RMSE of different estimators: (a) RMSE of position; (b) RMSE of velocity.}
\label{rmse}
\end{figure}

Fig.~\ref{rmse} shows the RMSEs of the position and velocity for the compared algorithms. It can be seen that the robust filters have a significant improvement except HRKF. Although both RCKF and LRKF are based on MCC, the proposed RCKF performs better than the LRKF, mainly because LRKF is based a linear regression, which introduces the linearzation error. The superiority of both RCKS and TCKS are also illustrated in Fig.~\ref{rmse} where RCKS has a similar performance as the TCKS.

\section{conclusion}
\label{con}

A new unified framework for robust Kalman filtering and smoothing based on the MCC has been proposed in this work. To ease the understanding of the proposed approach, we integrate the conventional cubature Kalman filter and smoother into our framework. The proposed robust Kalman filter and smoother is derived under the MCC, instead of the well-known MMSE, to cope with heavy-tailed noises in both the process and observation procedures. The intractable objective function based on MCC is then solved by the half-quadratic optimization method in an iterative manner. At each iteration the problem is reduced to a quadratic optimization problem, which can be efficiently solved by the conventional nonlinear Kalman filter and smoother. Numerical simulations demonstrate the superiority of the proposed robust filter and smoother when dealing with outliers in both states and measurements. The simulation results show that the proposed algorithms outperform state-of-art methods at comparable computational burden. One interesting future research direction is to investigate how to choose the kernel bandwidth of the MCC.

\appendix

\section{Cubature Kalman Filter and Smoother}
\label{A}
\numberwithin{equation}{section}
\setcounter{equation}{0}

We consider the SSM described in \eqref{SSM1} and \eqref{SSM2} without the contaminated noise (i.e., $\rho_1=\rho_2=0$) and the process and measurement noise follow
\begin{align*}
\bm v_{t-1}\thicksim\mathcal{N}(0,\bm Q_{t-1}),\quad
\bm w_{t}\thicksim\mathcal{N}(0,\bm R_{t}).
\end{align*}
The CKF has two steps involved, i.e., predicting and filtering. After the CKF procedure, the CKS is implemented by a backward smoothing step. The CKF and CKS are briefly summarized here for easy reference.
The algorithms are initialized with $\bm x_0\thicksim\mathcal{N}(\hat {\bm x}_{0|0},\bm P_{0|0})$ and the basic weighted cubature point set is given by $\{\bm \xi_{i},\omega_i\}$ for $i=1,\cdots,2n$, where $\bm \xi_{i}=\sqrt{n}[\bm I]_i$, $[\bm I]=[\bm I_{n},-\bm I_{n}]$ and $\omega_i=1/(2n)$.

\emph{Prediction:}
\begin{align}
\bm \chi_{i,t-1}=&f(\bm \xi_{i,t-1})\\
\bm \hat{\bm x}_{t|t-1}=&\sum_{i=1}^{2n}\omega_i\bm  \chi_{i,t-1}\\
\bm P_{t|t-1}=&\sum_{i=1}^{2n}\omega_i(\bm \chi_{i,t-1}-\bm \hat{\bm x}_{t|t-1})(\bm \chi_{i,t-1}-\bm \hat{\bm x}_{t|t-1})^T+\bm Q_{t-1}
\end{align}
where $\bm \xi_{i,t-1}$ is the transformed sigma point related to distribution $\mathcal{N}(\hat{\bm x}_{t-1|t-1},\bm P_{t-1|t-1})$, i.e.,
\begin{align}
\bm P_{t-1|t-1}&=\bm S_{t-1|t-1}\bm S_{t-1|t-1}^T\notag\\ \notag
\bm\xi_{i,t-1}&=\bm S_{t-1|t-1}\bm\xi_i+\hat{\bm x}_{t-1|t-1}
\end{align}

\emph{Filtering:}
\begin{align}
\hat{\bm x}_{t|t}&=\hat{\bm x}_{t|t-1}+\bm K_t(\bm y_t-\hat{\bm y}_t)\label{s_upt}\\
\bm P_{t|t}&=\bm P_{t|t-1}-\bm K_t\bm P_{yy}\bm K_t^T\\
\bm K_t&=\bm P_{xy}\bm P_{yy}^{-1}\label{G_upt}
\end{align}
where
\begin{align}
\bm \psi_{i,t}&=h(\bm \pi_{i,t}),\ \bm{\hat y}_t=\sum_{i=1}^{2n}\omega_i \bm \psi_{i,t} \\
\bm P_{yy}&=\sum_{i=1}^{2n}\omega_i\left(\bm \psi_{i,t}-\bm{\hat y}_t\right)\left(\bm \psi_{i,t}-\bm{\hat y}_t\right)^T+\bm R_t\label{I_upt} \\
\bm P_{xy}&=\sum_{i=1}^{2n}\omega_i\left(\bm \chi_{i,t}-\bm{\hat x}_{t|t-1}\right)\left(\bm \psi_{i,t}-\bm{\hat y}_t\right)^T
\end{align}
where $\bm \pi_{i,t}$ is the transformed sigma point related to distribution $\mathcal{N}(\hat{\bm x}_{t|t-1},\bm P_{t|t-1})$, i.e.,
\begin{align}
\bm P_{t|t-1}=\bm S_{t|t-1}\bm S_{t|t-1}^T,\ \bm \pi_{i,t}=\bm S_{t|t-1}\bm\xi_i+\bm \hat{\bm x}_{t|t-1} \notag
\end{align}

\emph{Smoothing}
\begin{align}
\hat{\bm x}_{t|T}&=\hat{\bm x}_{t|t}+\bm D_{t+1}(\hat{\bm x}_{t+1|T}-\hat{\bm x}_{t+1|t})\\
\bm P_{t|T}&=\bm P_{t|t}+\bm D_{t+1}(\bm P_{t+1|T}-\bm P_{t+1|t})\bm D_{t+1}^T
\end{align}
where
\begin{align}
\bm D_{t+1}&=\bm C_{t+1}\bm P_{t+1|t}^{-1}\\
\bm C_{t+1} &= \sum_{i=1}^{2n}\omega_i(\bm\xi_{i,t}-\hat{\bm x}_{t|t})(\bm \chi_{i,t}-\bm \hat{\bm x}_{t+1|t})^T
\end{align}
Here we should note that the filtered state is the same as the smoothed state at the end instant, so does their corresponding error covariance.

\section{Correntropy and Maximum Correntropy Criterion}
\label{B}
The correntropy, which is first proposed in~\cite{Liu:2007}, is a concept in information theoretic learning to deal with non-Gaussian noise. The correntropy is a generalized similarity measure between two arbitrary scalar random variables $X$ and $Y$, defined by
\begin{align}
V(X,Y)=E(\kappa_{\sigma}(X-Y))
\end{align}
where $E(\cdot)$ is the expectation and $\kappa(\cdot)$ is a kernel function with a control parameter $\sigma$, which satisfies Mercer's Theorem~\cite{Vapnik:1995}. The commonly used kernel, the Gaussian kernel, is considered in this paper, i.e.,
\begin{align}
\kappa_{\sigma}(X-Y)=\exp({-\frac{e^2}{2\sigma^2}})
\end{align}
where $e=X-Y$ is the difference between two variables, and $\sigma>0$ is the bandwidth of the Gaussian kernel.

Calculating the exact value of $V(X,Y)$ requires the joint distribution of $X$ and $Y$ which is barely known in practice. However, finite number of pairwise samples $\{x_i,y_i\}_{i=1}^N$ are often available. The correntropy can be estimated by
\begin{align}
V(x,y)\approx\frac{1}{N}\sum_{i=1}^N\kappa(x_i-y_i).
\end{align}

Correntropy has a number of nice properties that make it useful for non-Gaussian signal processing, especially in the impulsive noise environment. Correntropy is symmetric, positive and bounded. In addition, unlike the global similarity measure-mean square error (MSE) which only contains the second-order statistics, correntropy incorporates all even order moments~\cite{Liu:2007}. In geometric meaning, MSE gives the $\ell_2$ norm distance over $\{x,y\}$ while correntropy offers a hybird norm distance. Specifically, when two points are similar, correntropy behaves like the $\ell_2$ norm distance while it approaches the $\ell_1$ norm distance with increasing difference between two points. Finally, correntropy is equivalent to the $\ell_0$ norm distance if two points are far apart.

Based on the geometric meaning of correntropy, maximizing the correntropy of two different random variables can be used as a criterion in dealing with non-Gaussian noise problem, especially heavy-tailed noises, which leads to MCC. MCC has a close relationship with the M-estimator~\cite{Liu:2007}, and it is actually equivalent to the Welsch M-estimator~\cite{He:2014}. The superior performance of MCC in handling with the impulse noise has also been reported in~\cite{He:2014}. Hence MCC is a promising option to design robust filters and smoothers.

\section{Proof of Proposition~\ref{propo}}
\label{prop_proof}

In order to prove Proposition~\ref{propo}, we first introduce the following theorem from~\cite{charbonnier1997}:
\begin{theorem}
\label{t1}
Let $f(x)$ be a function that satisfies the following conditions:
\begin{enumerate}
  \item $f(x)\ge 0$ $\forall x$ with $f(0)=0$;
  \item $f(x)=f(-x)$;
  \item $f(x)$ continuously differentiable;
  \item $f'(x)\ge 0\  \forall x\ge 0$;
  \item $f'(x)/(2x)$ continuous and strictly decreasing on $[0,+\infty)$;
  \item $\lim\limits_{x\to+\infty}(f'(x)/(2x))=0$;
  \item $\lim\limits_{x\to 0^+}(f'(x)/(2x))=M$ where $0<M<+\infty$
\end{enumerate}
then
\begin{enumerate}
  \item there exists a strictly convex and decreasing function $g:\ (0,M]\to[0,\beta)$ where
  \begin{align}
  \beta = \lim_{x\to+\infty}\left(f(x)-x^2\frac{f'(x)}{2x}\right)
  \end{align}
  such that
  \begin{align}
  f(x)=\inf_{0<w\le M}\left(wx^2+g(w)\right)
  \end{align}
  \item for every fixed $\tilde{x}$, the value $w^{m}$ for which the minimum is reached, i.e., such that
      \begin{align}
      \inf_{0<w\le M}\left(wx^2+g(w)\right)=\left(w^{m}{\tilde x}^2+g(w^{m})\right)
      \end{align}
      is unique and given by
      \begin{align}
      w^{m}=\frac{f'(\tilde x)}{2\tilde x}
      \end{align}
\end{enumerate}
\end{theorem}
To use the above theorem to prove Proposition~\ref{propo}, let $f(x)=1-e^{-\frac{x^2}{2\sigma^2}}$. It is easy to check that $f(x)$ satisfies 1) to 7) in Theorem~\ref{t1}.  Therefore, we conclude that there exists a convex function $\theta(w):(0,M]\to [0,\beta)$ such that
\begin{align}
1-e^{-\frac{x^2}{2\sigma^2}}=\inf_{0<w\le M}\left(wx^2+\theta(w)\right)
\label{imp1}
\end{align}
with $M = 1/(2\sigma^2)$, $\beta=1$, and for a fixed $x$, the minimum is reached at $w=\frac{1}{2\sigma^2}e^{-\frac{x^2}{2\sigma^2}}$.

We can equivalently rewrite~\eqref{imp1} in terms of the supremum as
\begin{align}
e^{-\frac{x^2}{2\sigma^2}}=\sup_{0<w\le M}\left(-wx^2-\theta(w)+1\right)
\label{imp2}
\end{align}
Define $p=-2w\sigma^2$, and then~\eqref{imp2} is transformed as
\begin{align}
e^{-\frac{x^2}{2\sigma^2}}=\sup_{-1\le p<0}\left(\frac{p}{2\sigma^2}x^2-\theta(-\frac{p}{2\sigma^2})+1\right)
\end{align}
Furthermore, define $\psi(p)=\left(\theta(-\frac{p}{2\sigma^2})-1\right):[-1,0)\to [-1,0)$. We obtain
\begin{align}
\kappa_{\sigma}(x)=\sup_{-1\le p<0}\left(\frac{p}{2\sigma^2}x^2-\psi(p)\right)
\end{align}
and for a fixed $x$ the supremem is reached at
\begin{align}
p = -2\sigma^2\times\frac{1}{2\sigma^2}e^{-\frac{x^2}{2\sigma^2}}=-\kappa_{\sigma}(x)
\end{align}
Since $\psi(p)$ is obtained by an affine transformation from $\theta(w)$, $\psi(p)$ is a convex function due to the convexity invariance under affine maps.

\end{document}